\documentclass[journal]{IEEEtran}
\usepackage{color,soul}

\usepackage[fleqn]{amsmath}

\usepackage[hidelinks]{hyperref}

\usepackage{cite}

\ifCLASSINFOpdf
   \usepackage[pdftex]{graphicx}
\else
\fi
\usepackage{amsmath}

\usepackage{stfloats}

\begin{document}
%
\title{Parallel FDTD modelling \\of nonlocality in plasmonics}
%
%
%

\author{Joshua~Baxter,
        Antonino Cal\`a Lesina,
        and Lora Ramunno
\thanks{J. Baxter, A. Cal\`a Lesina, and L. Ramunno are with the Department of Physics and Center for Research in Photonics, University of Ottawa, Ottawa, Ontaria, Canada e-mail: jbaxt089@uottawa.ca.}
}

\maketitle

\begin{abstract}
As nanofabrication techniques become more precise, with ever smaller feature sizes, the ability to model nonlocal effects in plasmonics becomes increasingly important. While nonlocal models based on hydrodynamics
have been implemented using various computational electromagnetics techniques, the finite-difference time-domain (FDTD) version has remained elusive. Here we present a comprehensive FDTD implementation of nonlocal hydrodynamics, including for parallel computing. As a sub-nanometer step size is required to resolve nonlocal effects, a parallel implementation makes the computational cost of nonlocal FDTD more affordable. 
We first validate our algorithms for small spherical metallic particles, and find that nonlocality smears out staircasing artifacts at metal surfaces, increasing the accuracy over local models. We find this also for a larger nanostructure with sharp extrusions. The large size of this simulation, where nonlocal effects are clearly present, highlights the importance and impact of a parallel implementation in FDTD.  
\end{abstract}

\begin{IEEEkeywords}
FDTD, plasmonics, nonlocality, hydrodynamic plasma model, GNOR, parallel computing
\end{IEEEkeywords}

%
\IEEEpeerreviewmaketitle

\section{Introduction}
%
%
%
%
\IEEEPARstart{F}{abricating} objects with nanoscale precision is possible due to significant progress in nanofabrication techniques over the last few decades \cite{su_advances_2018}. As a result, plasmonic nanostructures and metamaterials \cite{schuller_plasmonics_2010} are having tremendous impact in many fields including biosensing \cite{mejia-salazar_plasmonic_2018}, quantum cryptography \cite{bozhevolnyi_plasmonics_2017}, nonlinear optics \cite{kauranen_nonlinear_2012}, photovoltaics \cite{enrichi_plasmonic_2018}, light emitting devices \cite{kim_localized_2015}, and precision medicine \cite{sharifi_plasmonic_2019}. 

Numerical modelling in plasmonics is typically based on optical models for bulk permittivity. These models -- such as the Drude model for free electron response, and the Lorentz and critical points models for bound electron response \cite{maier_plasmonics:_2007,etchegoin_analytic_2006} -- are based on the local response approximation (LRA), which assumes that the induced polarization or current at a given location depends only on the electromagnetic field at that same location. While appropriate for many applications, the Drude model is insufficient for modelling plasmonic structures smaller than 10 nm \cite{kreibig_optical_1985,raza_blueshift_2013,scholl_quantum_2012,baida_quantitative_2009}, as well as those containing sharp features or nanoscale gaps \cite{zhu_quantum_2016,ciraci_probing_2012}. The charge density near the surface of plasmonic objects is spread over a finite thickness on the order of the a few angstroms \cite{raza_nonlocal_2015,ciraci_probing_2012}, and thus for small features, cannot be treated as localized at the surface, as is implicit in LRA models.

%


The hydrodynamic plasma model treats the conduction band electrons as a free electron gas \cite{raza_unusual_2011,ciraci_probing_2012,kauranen_nonlinear_2012,raza_blueshift_2013,raza_nonlocal_2015}, accounting for free electron fluctuations via a pressure term. Unlike the Drude model, it does not make an LRA and thus nonlocality is incorporated. It has been found to correctly predict the expected spread of charge density near plasmonic surfaces as well as predict the blueshift in plasmon resonance with decreasing particle size \cite{scholl_quantum_2012,raza_blueshift_2013-1,raza_nonlocal_2015}. More recently, the effects of electron diffusion were incorporated within a generalized nonlocal optical response (GNOR) model, correctly predicting a broadened line-shape that was observed experimentally \cite{mortensen_generalized_2014}. 

Nonlocal hydrodynamic models have been implemented using several computational electromagnetic methods including the finite element method \cite{hiremath_numerical_2012}, the discontinuous Galerkin time domain method\cite{schmitt_3d_2017}, and the boundary element method \cite{zheng_boundary_2018}. A finite-difference time-domain (FDTD) implementation would be advantageous due to the wide-spread use of FDTD in the photonics community, its relative ease of implementation, its broadband capabilities, and its ability to produce time-domain movies. Only two attempts have been made \cite{mcmahon_topics_2011,zhang_fdtd_2016}, where one is based on an erroneous approximation, neither consider high performance computing and parallel implementations, and neither include electron diffusion. A correct and comprehensive FDTD implementation has remained unreported. 



In this paper, we present a parallel FDTD implementation of nonlocal hydrodynamics, including a GNOR implementation. A parallel implementation in FDTD is especially important for simulating nonlocality, as the grid cell size needs to be smaller than the Fermi wavelength $\lambda_F \sim 0.5$ nm to capture the spread of electron density \cite{gallinet_numerical_2015}. This can require a large amount of memory, and can present a prohibitive computational load. High-performance computing represents a viable solution. The FDTD rectangular-meshing scheme, where a field update at one location requires fields only in adjacent Yee-cells\cite{yee_numerical_1966}, lends itself well to parallel computing; indeed, message passing interface (MPI)-based FDTD solvers are well reported for LRA models \cite{lesina_convergence_2015}. Furthermore,  high-performance computing is becoming more accessible with cloud computing, computing consortiums, increased high-performance computing investments, and new, higher-level parallel programming languages such as Chapel\cite{noauthor_chapel:_nodate}.

The structure of this paper is as follows. We review nonlocal hydrodynamics in Section \ref{sec2}, including the most common version without electron diffusion, as well as GNOR, which does include electron diffusion. In Section \ref{sec3}, we derive from the nonlocal models FDTD update equations for the polarization field via the auxiliary differential equation (ADE) method.
In Section \ref{sec4}, we discuss the implementation of the nonlocal FDTD update equations for parallel computing within a MPI framework. In Section \ref{sec5}, we test our implementations by simulating the optical response of small metallic nanospheres and comparing our results to analytic solutions
and experimental results. We find an unexpected benefit of nonlocal versus LRA modelling: a marked decrease in staircasing artifacts at the metal boundary. Due to the rectangular discretization inherent in most FDTD approaches, fields can build up in an unphysical manner at plasmonic surfaces, and this can be particularly problematic in applications that rely on plasmonic near-field enhancement \cite{pernice_finite-difference_2010}. The incorporation of nonlocality significantly decreases this unphysical field build up. In Section \ref{sec6}, we simulate the response from a spherical nanoparticle containing sharp extrusions as a demonstration of a large-scale simulation that requires both parallel computing and nonlocal modelling. We find that despite the nanoparticle being larger, the sharp extrusions exhibit plasmonic features not accessible to the LRA. Finally, in Section \ref{sec7} we give concluding remarks. 

\section{Hydrodynamic models for nonlocality}\label{sec2}

The most general response of a linear optical material to incident radiation is described by

\begin{equation}
    \textbf{D}(\textbf{r},\omega)=\varepsilon_0\int\varepsilon(\textbf{r},\textbf{r}^{\prime},\omega)\textbf{E}(\textbf{r}^{\prime},\omega)d\textbf{r}^{\prime},
\end{equation}

\noindent where the dielectric function of the material, $\varepsilon(\textbf{r},\textbf{r}^{\prime},\omega)$, is nonlocal when the displacement field $\textbf{D}$ at one location depends on the electric field $\textbf{E}$ at other locations. In many applications, it is appropriate to use a local response approximation (LRA), wherein $\varepsilon(\textbf{r},\textbf{r}^{\prime},\omega) = \delta(\textbf{r}-\textbf{r}^{\prime})\varepsilon(\omega)$. In plasmonic modelling, the interaction of the electric field and the free electron plasma is typically described by the Drude model, which employs the LRA, where the dielectric function reduces to \cite{maier_plasmonics:_2007}

\begin{equation}
\varepsilon(\omega)=1-\frac{\omega_p^2}{\omega^2+i\gamma\omega},
\end{equation}

\noindent where $\omega_p$ is the plasma frequency, $\gamma$ is a collisional damping rate, and $i$ is the imaginary unit.

The hydrodynamic plasma model \cite{raza_nonlocal_2015} goes beyond the LRA, more accurately describing spatial-temporal free electron dynamics via

\begin{equation} \label{Hydro}
 \frac{\partial \textbf{v}}{\partial t} + (\textbf{v}\cdot\nabla)\textbf{v} = -\frac{e}{m}(\textbf{E}+\textbf{v}\times\textbf{B})-\gamma\textbf{v}-\frac{1}{m}\nabla\frac{\delta G[n]}{\delta n},
\end{equation}

\noindent where $\textbf{v}$ is the velocity field of the free electron plasma, $\textbf{E}$ and $\textbf{B}$ are the electric and magnetic fields, 
and $n$ is the free electron density. The energy functional $G[n]$ considers the internal kinetic energy of the electron gas and is usually taken to be the Thomas-Fermi functional, giving

\begin{equation}
    \frac{\delta G[n]}{\delta n}=\frac{h^2}{2m}\Big(\frac{3}{8\pi}n\Big)^{\frac{2}{3}},
\end{equation} 

\noindent where $h$ is Planck’s constant.  
As we are only considering the linear nonlocal response in this paper, we neglect the nonlinear terms (magnetic-Lorentz and convection) in Eq. \ref{Hydro} giving

\begin{equation} \label{LinHydro}
 \frac{\partial \textbf{v}}{\partial t} + \gamma\textbf{v} = -\frac{e}{m}\textbf{E} -\frac{\beta^2}{n}\nabla n,
\end{equation}

\noindent where $\beta^2=1/3 v_F^2$ and $v_F$ is the Fermi velocity. The free electron density fluctuations are accounted for in the last term of Eq. \ref{LinHydro}. This is often referred to as the pressure term, and is responsible for the known blue shift in the surface plasmon resonance with decreasing particle size \cite{scholl_quantum_2012,raza_nonlocal_2015}. Along with this formula for the velocity field, we require the continuity equation given by  

\begin{equation} \label{Cont}
	\frac{\partial n}{\partial t} = -\nabla \cdot(n\textbf{v}).
\end{equation} 

%

From here we consider two approaches. The first is the most widely used, and assumes a current density given by $\textbf{J}=-en\textbf{v}$. 
As we use a polarization field formulation in this paper, we set $\textbf{J}=\frac{\partial \textbf{P}_{NL}}{\partial t}$, where we have defined $\textbf{P}_{NL}$ to be the (nonlocal) free electron polarization field. The velocity field is thus $\textbf{v}=-\frac{1}{ne} \frac{\partial\textbf{P}_{NL}}{\partial t}$, and Eq. \ref{Cont} becomes

\begin{equation} \label{Nsol}
    n=n_0+\frac{1}{e}\nabla\cdot\textbf{P}_{NL},
\end{equation}

\noindent where $n_0$ is the equilibrium free electron density. From Eq. \ref{LinHydro}, we then obtain

\begin{equation}\label{NonlocDrude}
    \frac{\partial^2 \textbf{P}_{NL}}{\partial t^2} + \gamma \frac{\partial \textbf{P}_{NL}}{\partial t}=\varepsilon_0\omega_p^2\textbf{E}+\beta^2\nabla (\nabla\cdot\textbf{P}_{NL}),
\end{equation}

\noindent where $\omega_p=\sqrt{\frac{e^2n_0}{m\varepsilon_0}}$. We call Eq. \ref{NonlocDrude} the ``nonlocal Drude model" because it consists of the classical LRA Drude model plus one additional term that gives rise to nonlocality (\textit{i.e.}, the term proportional to $\beta^2$).

The model represented by Eq. \ref{NonlocDrude} differs from those used in the two previous nonlocal FDTD works. In Ref. \cite{mcmahon_topics_2011}, the gradient-divergence term was simplified to a Laplacian, and this resulted in spurious, non-physical resonances \cite{raza_unusual_2011}. In Ref. \cite{zheng_boundary_2018}, a current density formulation is used. A benefit to using a polarization field formulation is that it gives ready access to the free electron density via Gauss's Law, as we demonstrate in Section \ref{sec5}.

The second approach we consider includes electron diffusion, which was also found to play an important role in the optical response of small metallic particles \cite{mortensen_generalized_2014}. This has been described by the generalized nonlocal optical response (GNOR) model, where diffusion is considered by modifying the expression for the velocity field via $\textbf{v}=-\frac{1}{en} (\frac{\partial \textbf{P}_G}{\partial t}+\mathcal{D} \nabla(\nabla\cdot\textbf{P}_G))$, where $\mathcal{D}$ is the diffusion coefficient and where we have denoted the nonlocal GNOR free electron polarization field by $\textbf{P}_G$. GNOR correctly predicts both the blueshift in the plasmon resonance frequency as well as the broadening of the absorption peak with decreasing particle size. A thorough discussion of the nonlocal Drude and GNOR models is given in Ref. \cite{raza_nonlocal_2015}.

Using in Eqs. \ref{LinHydro} and \ref{Cont} the GNOR definition for the velocity field, we obtain the following time-domain nonlocal-diffusive hydrodynamics model for the polarization field, which we hereafter refer to as the ``GNOR" model:

\begin{equation}\label{GNOR}
    \frac{\partial^2 \textbf{P}_G}{\partial t^2} + \gamma \frac{\partial \textbf{P}_G}{\partial t}=\varepsilon_0\omega_p^2\textbf{E}+\eta\nabla (\nabla\cdot\textbf{P}_G) + \mathcal{D}\frac{\partial}{\partial t}\nabla (\nabla\cdot\textbf{P}_G), 
\end{equation}

\noindent where $\eta=\beta^2+\mathcal{D}\gamma$. 

It is from Eqs. \ref{NonlocDrude} and \ref{GNOR} that we derive in the next section our FDTD update equations for implementing the two different models of the free electron response of a plasmonic material, one accounting for nonlocality only (Eq. \ref{NonlocDrude}), and the other nonlocality with diffusion (Eq. \ref{GNOR}).


To properly model the optical response of many plasmonic materials, one must also include the contribution of bound electrons. We employ the LRA-based \textit{N}-critical points model that assumes a susceptibility of the form
\begin{multline}
\label{CritPoint}
    \chi_{CP}(\omega)=(\varepsilon_\infty - 1) \\ + \sum_{p=1}^{N} A_p\Omega_p\Big( \frac{e^{i\phi_p}}{\Omega_p-\omega-i\Gamma_p} + \frac{e^{-i\phi_p}}{\Omega_p+\omega+i\Gamma_p}\Big),
\end{multline}

\noindent where $\varepsilon_\infty$ is the infinite frequency permittivity, and $A_p,\Omega_p,\Gamma_p$, and $\phi_p$ are fitting parameters. This model can be readily transformed to the time-domain for FDTD implementation, as described in detail in Ref. \cite{prokopidis_unified_2013}.

\section{Update equations for Nonlocal FDTD}\label{sec3}

\begin{table}[b!]
\centering
\caption{Yee Cell positions of the electric, magnetic, and polarization fields}
\begin{tabular}{|p{0.7in}||p{0.5in}||p{0.7in}||p{0.5in}|} \hline 
 & \textbf{Electric } & \textbf{Magnetic } & \textbf{Polarization} \\ \hline 
$E_x,H_x,P_x$ & i+1/2, j, k & i, j+1/2, k+1/2 & i+1/2, j, k \\ \hline 
$E_y,H_y,P_y$ & i, j+1/2, k & i+1/2, j, k+1/2 & i, j+1/2, k \\ \hline 
$E_z,H_z,P_z$ & i, j, k+1/2 & i+1/2, j+1/2, k & i, j, k+1/2 \\ \hline 
\end{tabular}\label{Tab:yeetable}
\end{table}

We derive in this section the FDTD update equations for the time-domain nonlocal Drude and GNOR models (Eqs. \ref{NonlocDrude} and \ref{GNOR}, respectively), using the ADE method. We discretize our domain via the Yee cell \cite{yee_numerical_1966} where the electric fields are collocated with the polarization fields in time and space; the Yee cell positions of the electric, magnetic, and polarization fields used in this paper are listed in Table \ref{Tab:yeetable}. We denote the free electron polarization field by $\textbf{P}_f$ (which signifies either $\textbf{P}_{NL}$ or $\textbf{P}_G$), and the bound electron polarization field by $\textbf{P}_{CP}$.

The FDTD update algorithm at time = $n\Delta t$, where $\Delta t$ is the time step size, consists of updating (in order) the:
\vskip 0.05in
(1) magnetic field $\textbf{H}^{n+1/2}=f_1 (\textbf{H}^{n-1/2},\textbf{E}^n )$, 

(2) electric field $\textbf{E}^{n+1}=f_2 (\textbf{E}^n,\textbf{P}_f^n,\textbf{P}_{CP}^n,\textbf{H}^{n+1/2})$,  

(3) bound charge polarization $\textbf{P}_{CP}^{n+1}=f_3(\textbf{E}^{n+1},\textbf{P}_{CP}^n)$,

(4) free charge polarization $\textbf{P}_f^{n+1}=f_4 (\textbf{E}^{n+1},\textbf{P}_f^n)$,
\vskip 0.05in
\noindent via update equations  $f_1$, $f_2$, $f_3$, and $f_4$, whose form and required inputs depend on the model from which they are derived. The equation $f_1$ is the magnetic field update derived through discretization of the Maxwell-Faraday equation\cite{taflove_computational_2005}; we do not derive this here because in plasmonic simulations it is typically unchanged from the vacuum equation. In what follows, we  present the electric field update equation $f_2$ (derived from the Maxwell-Amp\`ere law), the bound charge polarization update equation $f_3$ (derived from Eq. \ref{CritPoint}), and the free charge polarization update $f_4$ (derived from Eq. \ref{NonlocDrude} for the nonlocal Drude model, and Eq. \ref{GNOR} for the GNOR model). Though we present $f_4$ for two nonlocal models, one must chose which to use  -- they cannot be used simultaneously. 

We start by deriving $f_4$ from the nonlocal Drude model. Using central differencing, we discretize Eq. \ref{NonlocDrude} centered at time $n\Delta t$, to obtain 

\begin{multline}
    \frac{\textbf{P}_{NL}^{n+1}-2\textbf{P}_{NL}^{n}+\textbf{P}_{NL}^{n-1}}{\Delta t^2} + \gamma\frac{\textbf{P}_{NL}^{n+1}-\textbf{P}_{NL}^{n-1}}{2\Delta t} 
    \\ =\varepsilon_0\omega_p^2\textbf{E}^n  +\beta^2\nabla (\nabla\cdot\textbf{P}_{NL}^n).
\end{multline}

\noindent Further, using weighted central averaging as discussed in Ref. \cite{pernice_finite-difference_2010}, we set $\textbf{E}^n = (\textbf{E}^{n-1}+2\textbf{E}^n+\textbf{E}^{n+1})/4$ to obtain

\begin{align}\label{NonlocalUpdate}
\textbf{P}_{NL}^{n+1}&=D_1\textbf{P}_{NL}^{n}+D_2\textbf{P}_{NL}^{n-1}+D_3(\textbf{E}^{n-1}+2\textbf{E}^n+\textbf{E}^{n+1})\nonumber \\ &+D_{NL}\nabla(\nabla\cdot\textbf{P}_{NL}^n),
\end{align}

\noindent where



\begin{align} 
    &D_1 = \frac{2}{D_d\Delta t^2} \label{d1},\\
    &D_2 = \frac{1}{D_d}\Bigg(\frac{\gamma}{2\Delta t} - \frac{1}{\Delta t^2} \Bigg)  \label{d2},\\
    &D_3 = \frac{\varepsilon_0\omega_p^2}{4D_d} \label{d3}, \\
    &D_{NL}=\frac{\beta^2}{D_d} \label{d4}, \\
    &D_d = \Bigg(\frac{\gamma}{2\Delta t} + \frac{1}{\Delta t^2} \Bigg) \label{dd}.
\end{align}

\noindent As the $x$, $y$, and $z$ components of $\nabla(\nabla\cdot\textbf{P}_{NL}^n)$ must be collocated with the $x$, $y$, and $z$ components of $\textbf{P}_{NL}$, we have

\begin{subequations}\label{GradDiv}
\begin{multline}
    \nabla(\nabla\cdot\textbf{P}_{NL}^n)\Bigg|^{i+1/2,j,k}_x = \frac{\partial}{\partial x}\Bigg( \frac{\partial P_x}{\partial x} + \frac{\partial P_y}{\partial y} + \frac{\partial P_z}{\partial z} \Bigg)\Bigg|^{i+1/2,j,k}_x \\
    \approx \frac{P_x^{i+3/2,j,k} - 2P_x^{i+1/2,j,k} + P_x^{i-1/2,j,k}}{\Delta x^2}\\ + \frac{P_y^{i+1,j+1/2,k}-P_y^{i+1,j-1/2,k}-P_y^{i,j+1/2,k}+P_y^{i,j-1/2,k}}{\Delta x \Delta y} \\
    +\frac{P_z^{i+1,j,k+1/2}-P_z^{i+1,j,k-1/2}-P_z^{i,j,k+1/2}+P_z^{i,j,k-1/2}}{\Delta x \Delta z},
\end{multline}

\begin{multline}
    \nabla(\nabla\cdot\textbf{P}_{NL}^n)\Bigg|^{i,j+1/2,k}_y = \frac{\partial}{\partial y}\Bigg( \frac{\partial P_x}{\partial x} + \frac{\partial P_y}{\partial y} + \frac{\partial P_z}{\partial z} \Bigg)\Bigg|^{i,j+1/2,k}_y \\
    \approx \frac{P_y^{i,j+3/2,k} - 2P_y^{i,j+1/2,k} + P_y^{i,j-1/2,k}}{\Delta y^2}\\ + \frac{P_x^{i+1/2,j+1,k}-P_x^{i-1/2,j+1,k}-P_x^{i+1/2,j,k}+P_x^{i-1/2,j,k}}{\Delta x \Delta y} \\
    +\frac{P_z^{i,j+1,k+1/2}-P_z^{i,j+1,k-1/2}-P_z^{i,j,k+1/2}+P_z^{i,j,k-1/2}}{\Delta y \Delta z},
\end{multline}

\begin{multline}
    \nabla(\nabla\cdot\textbf{P}_{NL}^n)\Bigg|^{i,j,k+1/2}_z = \frac{\partial}{\partial z}\Bigg( \frac{\partial P_x}{\partial x} + \frac{\partial P_y}{\partial y} + \frac{\partial P_z}{\partial z} \Bigg)\Bigg|^{i,j,k+1/2}_z \\
    \approx \frac{P_z^{i,j,k+3/2} - 2P_z^{i,j,k+1/2} + P_z^{i,j,k-1/2}}{\Delta z^2}\\ + \frac{P_x^{i+1/2,j,k+1}-P_x^{i-1/2,j,k+1}-P_x^{i+1/2,j,k}+P_x^{i-1/2,j,k}}{\Delta x \Delta z} \\
    +\frac{P_y^{i,j+1/2,k+1}-P_y^{i,j-1/2,k+1}-P_y^{i,j+1/2,k}+P_y^{i,j-1/2,k}}{\Delta y \Delta z},
\end{multline}
\end{subequations}

\noindent where, for brevity, the components of $\textbf{P}_{NL}$ are written as $P_a$ for $a=x,y,z$.

For the GNOR model, the $f_4$ update equation includes Eqs. \ref{NonlocalUpdate}-\ref{GradDiv} (with $\beta^2$ replaced by $\eta$) along with a suitable discretization for the last term of Eq. \ref{GNOR}, the only term that does not appear in the nonlocal Drude model. As this term cannot be achieved via central differencing, we use an alternate second order finite difference scheme given by 

\begin{equation}
    \frac{\partial}{\partial t}\nabla (\nabla\cdot\textbf{P}_G^n) \approx \frac{3\nabla (\nabla\cdot\textbf{P}_G^n)-4\nabla (\nabla\cdot\textbf{P}_G^{n-1})+\nabla (\nabla\cdot\textbf{P}_G^{n-2})}{2\Delta t}.
\end{equation}

\noindent The GNOR $f_4$ update then becomes

\begin{multline}\label{GNORUpdate}
\textbf{P}_G^{n+1}=D_1\textbf{P}_G^{n}+D_2\textbf{P}_G^{n-1}+D_3(\textbf{E}^{n-1}+2\textbf{E}^n+\textbf{E}^{n+1})\\+D_{G1}\nabla(\nabla\cdot\textbf{P}_G^n) \\
+D_{G2}(3\nabla (\nabla\cdot\textbf{P}_G^n)-4\nabla (\nabla\cdot\textbf{P}_G^{n-1})+\nabla (\nabla\cdot\textbf{P}_G^{n-2})),
\end{multline}

where
{\setlength{\mathindent}{0.7cm}
\begin{align}
    &D_{G1} =\frac{\eta}{D_d},\\
    &D_{G2} =\frac{\mathcal{D}}{2\Delta t D_d}.
\end{align}}

The $f_3$ update equation for the bound charge density is derived from the critical points model (Eq. \ref{CritPoint}) in Ref. \cite{prokopidis_unified_2013}. We state it here:

\begin{equation}\label{critpointupdate}
    \textbf{P}_{CP,p}^{n+1}=C_{1p}\textbf{P}_{CP,p}^n+C_{2p}\textbf{P}_{CP,p}^{n-1}+C_{3p}\textbf{E}^{n+1}+C_{4p}\textbf{E}^n+{C_{5p}\textbf{E}}^{n-1}
\end{equation}
\noindent where 
\begin{align}
&C_{1p}=\frac{1}{C_p}\left(\frac{2}{\Delta t^2}-\frac{\Omega_p^2+\Gamma_p^2}{2}\right)\\
&C_{2p}=\frac{1}{C_p}\left(\frac{\Gamma_p}{\Delta t}-\frac{1}{\Delta t^2}-\frac{\Omega_p^2+\Gamma_p^2}{4}\right)\\
&C_{3p}=\frac{C_{4p}}{2}-\frac{\varepsilon_0A_p\Omega_p\sin{\phi_p}}{2\Delta t C_p}\\
&C_{4p}=\frac{\varepsilon_0A_p\Omega_p\left(\Omega_p\cos{\phi_p}-\Gamma_p\sin{\phi_p}\right)}{C_p}\\
&C_{5p}=\frac{C_{4p}}{2}+\frac{\varepsilon_0A_p\Omega_p\sin{\phi_p}}{2\Delta t C_p} \\
&C_p = \frac{\Gamma_p}{\Delta t} + \frac{1}{\Delta t^2} + \frac{\Omega_p^2+\Gamma_p^2}{4},
\end{align}

\noindent where $p$ denotes the $p^{th}$ critical point. 

Next we turn to the $f_2$ update equation for $\textbf{E}^{n+1}$. Ampère's law is discretized in time to give 

\begin{multline}\label{discreteamperes}
    \varepsilon_0\varepsilon_\infty\frac{\textbf{E}^{n+1}-\textbf{E}^n}{\Delta t} + \frac{\textbf{P}_f^{n+1}-\textbf{P}_f^n}{\Delta t} + \sum_{p=1}^N\frac{(\textbf{P}_{CP,p}^{n+1}-\textbf{P}_{CP,p}^n)}{\Delta t}\\=\nabla\times\textbf{H}^{n+1/2}.
\end{multline}

\noindent For the nonlocal Drude model we set $\textbf{P}_f=\textbf{P}_{NL}$, and plug Eq. \ref{NonlocalUpdate} into Eq. \ref{discreteamperes} to obtain


{\setlength{\mathindent}{0.7cm}

\begin{equation}
\begin{split}
&\textbf{E}^{n+1}\left(\varepsilon_0\varepsilon_\infty+D_3+\sum_{p=1}^{N}C_{3p}\right)\\ &= \Delta t(\nabla\times\textbf{H}^{n+1/2}) +\left(\varepsilon_0\varepsilon_\infty-2D_3+\sum_{p=1}^{N}C_{4p}\right) \textbf{E}^n \\ & - \left(D_3 + \sum_{p=1}^{N}C_{5p}\right)\textbf{E}^{n-1} -(D_1 - 1)\textbf{P}_{NL}^n \\ &- \sum_{p=1}^{N}(C_{1p} - 1)\textbf{P}_{CP,p}^n - D_2\textbf{P}_{NL}^{n-1} \\ &-\sum_{p=1}^{N}C_{2p}\textbf{P}_{CP,p}^{n-1}-D_{NL}\nabla(\nabla\cdot\textbf{P}_{NL}^n).
\end{split}
\end{equation}

\noindent For the GNOR model we set $\textbf{P}_f=\textbf{P}_{G}$, and plug Eq. \ref{GNORUpdate} into Eq. \ref{discreteamperes} to obtain


\begin{equation}
\begin{split}
&\textbf{E}^{n+1}\left(\varepsilon_0\varepsilon_\infty+D_3+\sum_{p=1}^{N}C_{3p}\right)\\ &= \Delta t(\nabla\times\textbf{H}^{n+1/2})+\left(\varepsilon_0\varepsilon_\infty-2D_3+\sum_{p=1}^{N}C_{4p}\right)\textbf{E}^n \\ &- \left(D_3 + \sum_{p=1}^{N}C_{5p}\right)\textbf{E}^{n-1} 
    -(D_1 - 1)\textbf{P}_G^n \\& - \sum_{p=1}^{N}(C_{1p} - 1)\textbf{P}_{CP,p}^n - D_2\textbf{P}_G^{n-1} \\ &-\sum_{p=1}^{N}C_{2p}\textbf{P}_{CP,p}^{n-1} -D_{G1}\nabla(\nabla\cdot\textbf{P}_G^n)\\ &
    -D_{G2}(3\nabla(\nabla\cdot\textbf{P}_G^n)-4\nabla(\nabla\cdot\textbf{P}_G^{n-1})+\nabla(\nabla\cdot\textbf{P}_G^{n-2})).
\end{split}
\end{equation}

{\setlength{\mathindent}{0.7cm}

\noindent For the GNOR update, to avoid recalculation of $\nabla\left(\nabla\cdot\textbf{P}_{G}^{n-1}\right)$, and $\nabla\left(\nabla\cdot\textbf{P}_{G}^{n-2}\right)$, we store them in arrays; therefore GNOR requires additional memory.

%


The update equations derived above must be implemented in all cells in which the plasmonic material exists. However, care needs to be taken if the nonlocal material extends to the boundary of the simulation domain. Terminating a nonlocal material with a perfectly matched layer (PML) may result in instability and convergence issues. Within a total field/scattered field (TF/SF) framework  -- which is applicable to many plasmonic nanostructure scattering problems -- this can be overcome by using the nonlocal model only in the total field region. If TF/SF cannot be used, such as for geometries that include a plasmonic substrate, one may use the LRA-Drude model in the PML region, and the nonlocal models everywhere else. 

Finally, we would like to highlight that since the polarization fields are nonlocal, an additional boundary condition is required at the interface between the plasmonic and external media. In our FDTD implementation, we impose the Pekar additional boundary condition \cite{halevi_generalised_1984} by setting $\textbf{P}_{f}=0$ outside of the plasmonic structure (that is, by not updating $\textbf{P}_{f}$ outside the structure).

\section{Parallel implementation of nonlocal FDTD}\label{sec4}

In this section we describe a parallel FDTD scheme using the message passing interface (MPI) framework. Nonlocal simulations require a lot of memory since the step-size $\Delta x$ needs to be smaller than the Fermi wavelength. In turn, significant computation time is required because the Courant – Friedrichs – Lewy condition restricts the time-step $\Delta t$ according to $\Delta x$. Parallel computing thus becomes essential.  

The simulation domain is decomposed into $n_x\times n_y\times n_z$ MPI processes where $n_d$ is the number of MPI processes in the $d$ direction, with $d=x,y,z$. Each MPI process is identified by a vector $(m_x,m_y,m_z)$ which gives its relative spatial position within the simulation domain, where $0 \leq m_d < n_d$. 

Each process performs field updates within its own subdomain ({\it i.e.}, its section of the simulation domain) defined according to a local grid with $(N_x+1)\times(N_y+1)\times(N_z+1)$ points. The vector $(i,j,k)$ identifies an individual grid cell within the local grid, where $i$ ranges from $0$ to $N_x$, $j$ from $0$ to $N_y$, and $k$ from $0$ to $N_z$, inclusively. The electric and magnetic field components at $(i,j,k)$ correspond to different locations in physical space within the grid cell, according to the Yee cell shown in Table \ref{Tab:yeetable}. For example, $E_x(i,j,k)$ refers to $E_x(i\Delta x + \Delta x/2,j\Delta y,k\Delta z)$, whereas $E_y(i,j,k)$ refers to  $E_y(i\Delta x,j\Delta y + \Delta y/2,k\Delta z)$.  

The use of domain decomposition requires that data from the boundaries of subdomains be transferred to other subdomains at each time step. For subsequent updates to be executed efficiently, an overlap of information is required between adjacent subdomains. For example, the cells $(N_x,j,k)$ in subdomain $(m_x,m_y,m_z)$ represents the same physical locations as the cells $(0,j,k)$ in subdomain $(m_x+1,m_y,m_z)$. The update scheme we describe below guarantees that the update equations for a given field component at a given physical location are only applied once. 


\begin{table}[!t]
\centering
\caption{Subdomain update scheme for different field components.}
\begin{tabular}{|p{0.55in}||p{0.7in}||p{0.7in}||p{0.7in}|} \hline 
Component & $x$ dimension & $y$ dimension & $z$ dimension \\ \hline 
$E_x$ and $P_x$ & $0\to N_x-1$ & $1\to N_y$ & $1\to N_z$ \\ \hline 
$E_y$ and $P_y$ & $1\to N_x$ & $0\to N_y-1$ & $1\to N_z$ \\ \hline 
$E_z$ and $P_z$ & $1\to N_x$ & $1\to N_y$   & $0\to N_z-1$ \\ \hline 
$H_x$ & $1\to N_x$ & $0\to N_y-1$ & $0\to N_z-1$ \\ \hline 
$H_y$ & $0\to N_x-1$ & $1\to N_y$ & $0\to N_z-1$ \\ \hline 
$H_z$ & $0\to N_x-1$ & $0\to N_y-1$ & $1\to N_z$ \\ \hline 

\end{tabular}\label{EHexchange}
\end{table}

\begin{figure}[!t]
\centering
\includegraphics[width=\linewidth]{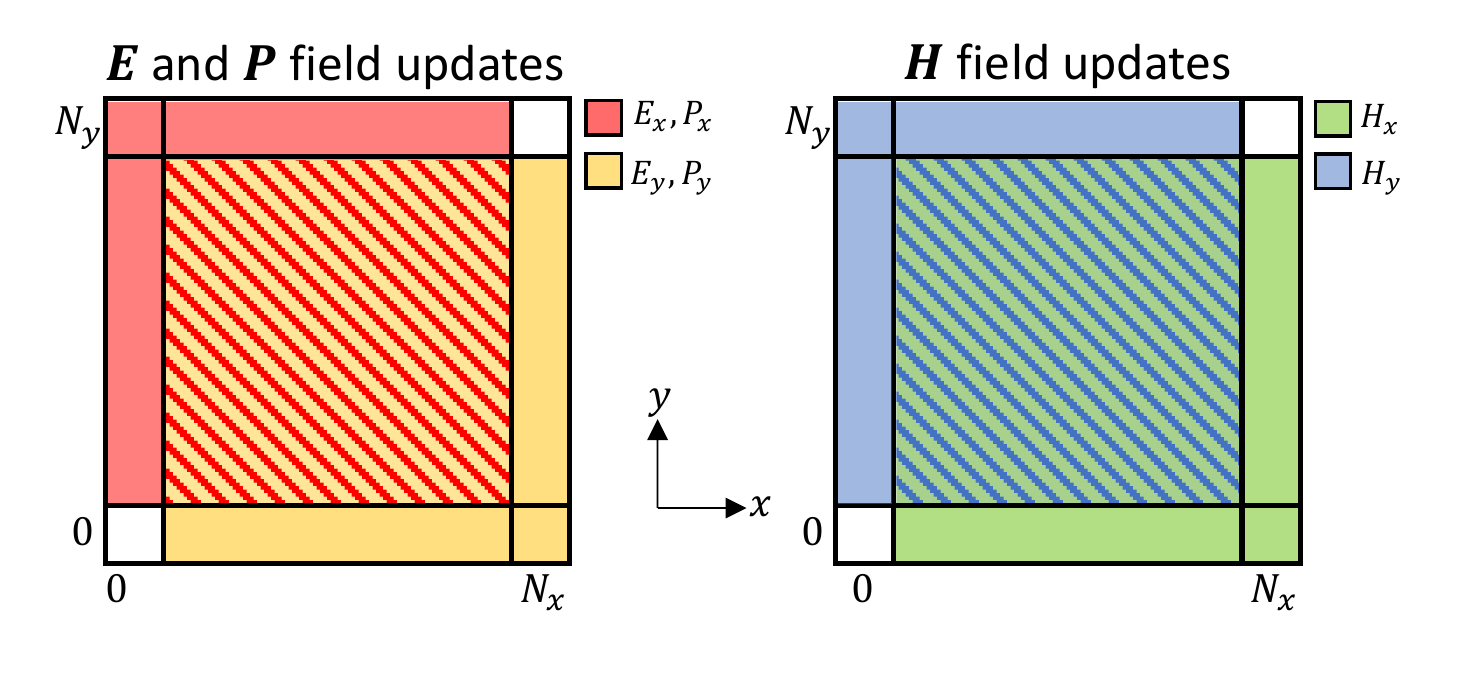}
\caption{Two dimensional representation of the field update regions within a subdomain for the electric/polarization fields (left) and the magnetic fields (right). Different colours are chosen to represent the different field components. The hatched regions indicate where both field components are updated.}
\label{fig:FieldUpdates}
\end{figure}

The components of the magnetic and electric fields in a subdomain are updated via $f_1$ and $f_2$, respectively, however each is updated for different ranges of indices $(i,j,k)$ as summarized in Table \ref{EHexchange}. The notation $0\to N_d$ means we update from index $0$ to index $N_d$, inclusively. Bound and free charge polarization fields are updated via $f_3$ and $f_4$, respectively, for the same ranges of indices as for the electric field, as listed in Table \ref{EHexchange}. To understand this visually, the field update regions for the $x$ and $y$ components of all fields within a single subdomain are illustrated in Fig. \ref{fig:FieldUpdates}.

While the components of the magnetic, electric, and bound charge polarization fields in a subdomain all require $(N_x+1)\times(N_y+1)\times(N_z+1)$ values to be stored,  the free charge polarization fields requires $(N_x+2)\times(N_y+2)\times(N_z+2)$. This is because the free charge polarization field updates via Eq. \ref{GradDiv} requires information from additional cells. For example, to update $P_{x,NL}(0,j,k)$ via Eq. \ref{GradDiv} (a), we need $P_{x,NL}(-1,j,k)$; in general, we need an extra cell in each dimension to store the $-1$ index. However, we do not calculate updates at this index, as $(-1,j,k)$ in subdomain $(m_x,m_y,m_z)$ obtain their values from a transfer of data from cells $(N_x-1,j,k)$ in subdomain $(m_x-1,m_y,m_z)$.

\begin{figure}[!t]
\includegraphics[width=3in]{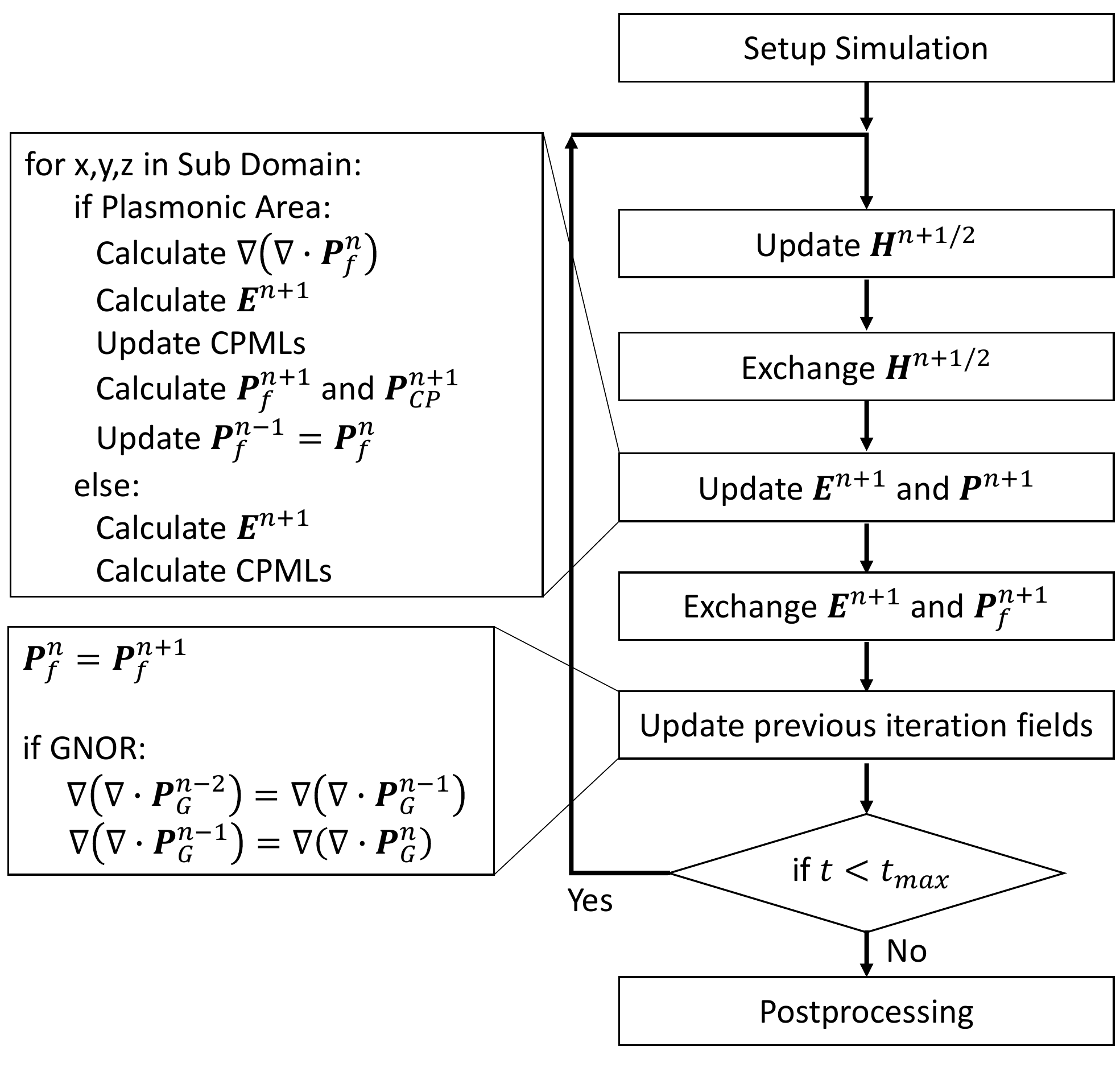}
\caption{FDTD algorithm for the nonlocal Drude and GNOR models in an MPI framework.}
\label{fig:Alg}
\end{figure}

We now discuss the flow of the FDTD algorithm for the both the nonlocal Drude and GNOR models, including what data needs to be communicated, and when. This is summarized in Fig. \ref{fig:Alg}, and we go through each step in detail in the following.
 
The first update to execute after the setup of the simulation is that of the magnetic field via $f_1$. Once this has been completed in each subdomain, a subset of the magnetic field values at the subdomain boundaries must be communicated to adjacent subdomains. The necessary communications for $H_x$ are given in the first row of Table \ref{Allexchange}; those for the other components of $\textbf{H}$ can be obtained from this table by an ordered permutation of $x\to y\to z$. The inter-subdomain communications in the $x$ and $y$ directions are illustrated in the top part of Fig. \ref{fig:FieldExchange}. Note that the data transfers of the magnetic fields are all made in the ``backward'' direction. The inset in Fig. \ref{fig:FieldExchange} details all data transfers along $x$ and their relative Yee cell positions. For example, $H_y(0,j,k)$ updated locally in subdomain $(m_x+1,m_y,m_z)$ is passed to $H_y(N_x,j,k)$ in subdomain $(m_x,m_y,m_z)$.

After the magnetic field data transfer, the electric field is updated in each subdomain via $f_2$, after which the bound and free charge polarization fields are updated in each subdomain via $f_3$ and $f_4$, respectively. Subsequently, a subset of their values at the subdomain boundaries need to be communicated to adjacent subdomains. The necessary communications for $E_x$ are given in the second row of Table \ref{Allexchange}; again, those for the other components are obtained by an ordered permutation of  $x\to y\to z$. The inter-subdomain communications electric field in the $x$ and $y$ directions are also illustrated in the top part of Fig. \ref{fig:FieldExchange}, with further detail given in the inset. Unlike the magnetic field data transfers, those for the electric field are made in the ``forward'' direction. For example, $E_y(N_x,j,k)$ updated in subdomain $(m_x,m_y,m_z)$, is passed to $E_y(0,j,k)$ in subdomain $(m_x+1,m_y,m_z)$.
Note that only the $\textbf{E}$ and $\textbf{H}$ components tangential to the direction of the data transfer need to be exchanged. 

\begin{figure}[!t]
\centering
\includegraphics[width=\linewidth]{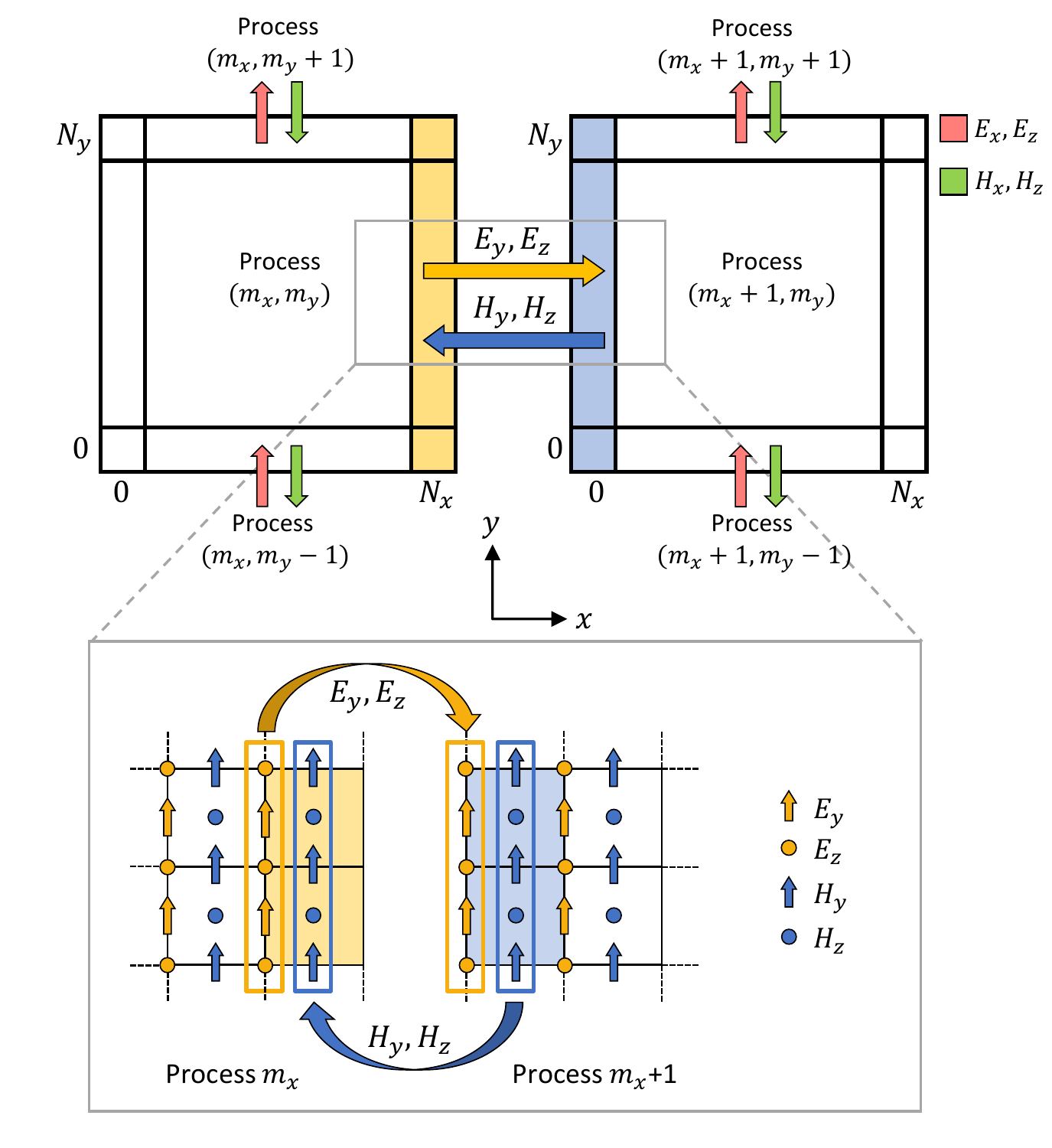}
\caption{Two dimensional representation of the magnetic and electric field data transfers between adjacent subdomains. The different colours represent the different field components as indicated.  The inset illustrates all field component transfers along $x$, and their relative positions in the Yee cell.}
\label{fig:FieldExchange}
\end{figure}

\begin{figure}[!t]
\centering
\includegraphics[width=\linewidth]{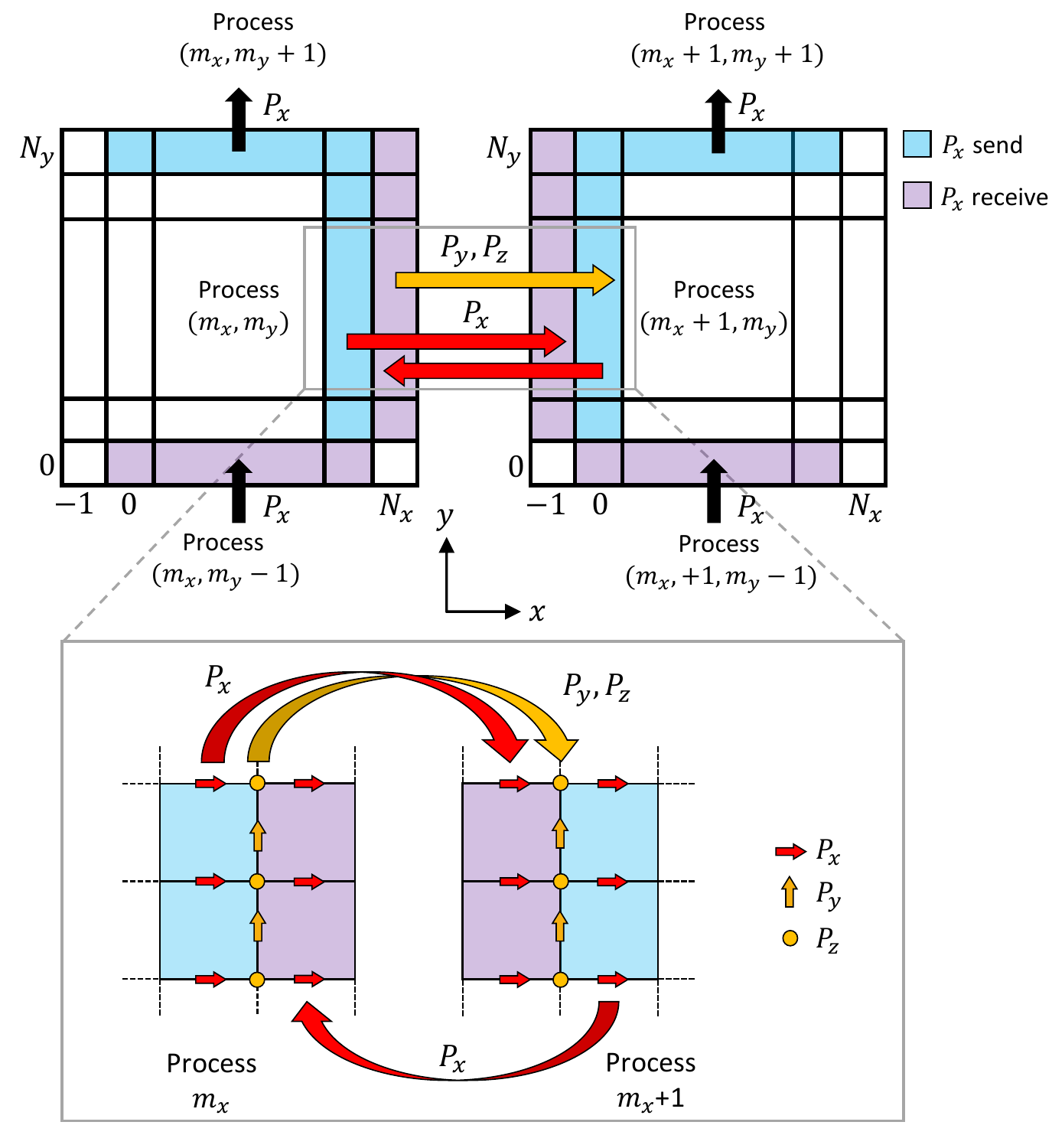}
\caption{Two dimensional representation of the data transfers between adjacent subdomains for the $x$ component of the free charge polarization field. The $j=-1$ row in the $y$ direction is not shown or required for $P_x$, but would be required for $P_y$. The blue shade represents regions where $P_x$ is locally calculated and sent to adjacent processes, and the purple represents regions where $P_x$ is received from adjacent processes. The inset illustrates all free charge polarization field component transfers along $x$, and their relative positions in the Yee cell.}
\label{fig:FieldExchangePx}
\end{figure}

In general, FDTD updates for LRA polarization models only require the collocated electric and polarization field values so that the polarization field values need not be communicated to other subdomains. Thus no communication is necessary for the bound charge polarization field $\textbf{P}_{CP}$. This is not true for the nonlocal models. The necessary communications for the $x$ component of $\textbf{P}_f$ are given in the third row of Table \ref{Allexchange} and visualized in the top part Fig. \ref{fig:FieldExchangePx}. The communication for the $y$ and $z$ field components are obtained from Table \ref{Allexchange} again via an ordered permutation of $x\to y\to z$. The inset of Fig. \ref{fig:FieldExchangePx} shows the data transfers along $x$ for all field components. Note that there are now four data transfers for every $\textbf{P}_f$ component -- three ``forward'' and one ``backward'' -- due to the increased data required for updates via Eq. \ref{GradDiv}. Thus, the inclusion of nonlocality doubles the number of communications required at each time-step. This has an effect on performance and scalability and is discussed further in Section \ref{sec5}.

It is worth noting that MPI is not the only solution for parallel computing as new higher-level languages are being introduced for this purpose. Chapel, a language produced by Cray \cite{noauthor_chapel:_nodate}, allows for algorithm implemenation on a distributed system without the challenges of MPI. For example, Ref. \cite{barrett_finite_nodate} presents a finite difference implementation of Poisson’s equation in Chapel. One may also use a shared memory implementation where the memory is shared amongst the processes (or threads) and therefore no data need be communicated. This can be implemented via OpenMP \cite{noauthor_openmp_nodate} for multi-threading on CPUs or via CUDA \cite{noauthor_cuda_2014} or OpenCL \cite{noauthor_opencl_2013} on graphics processing units (GPUs). Indeed, GPU-FDTD implementations are well reported in literature\cite{demir_compute_2010}. Since GPUs can launch thousands of parallel threads that all have access to shared memory, a GPU-based implementation of nonlocal FDTD requires no special treatment beyond what was presented in Section \ref{sec3}. While GPUs do suffer from memory constraints, they can be still be useful for smaller nonlocal plasmonic simulations. 

\begin{table}[!h]
\centering
\caption{Data transfer protocol for the $x$-component of the fields.}
\begin{tabular}{|p{0.15in}||p{2.9in}|} \hline 
$H_x$ & 	Transfer $y=0$ plane backward to adjacent $y=N_y$ plane\newline 
		   	Transfer $z=0$ plane backward to adjacent $z=N_z$ plane\\ \hline 
$E_x$ & 	Transfer $y=N_y$ plane forward to adjacent $y=0$ plane\newline 
			Transfer $z=N_z$ plane forward to adjacent $z=0$ plane\\ \hline 
$P_{f,x}$ & Transfer $x=0$ plane backward to adjacent $x=N_x$ plane\newline
			Transfer $x=N_x-1$ plane forward to adjacent $x=-1$ plane\newline 
			Transfer $y=N_y$ plane forward to adjacent $y=0$ plane\newline
			Transfer $z=N_z$ plane forward to adjacent $z=0$ plane\\ \hline 
\end{tabular}\label{Allexchange}
\end{table}

\section{Nonlocal FDTD applied to small spheres}\label{sec5}

In this section, we test and validate our FDTD implementations of the nonlocal models by using them to simulate the optical response of small silver nanospheres. We compare our results to (1) analytic solutions based on Mie theory, (2) LRA FDTD plasmonic simulations that employ the LRA Drude model for free-electron response, and (3) experimental results from the literature. Not only do our nonlocal FDTD simulations agree well with analytic and experimental results, we also find an unexpected benefit over the LRA approach: a pronounced reduction of staircasing artifacts.




\begin{figure}[t!]
\centering
\includegraphics[width=0.8\linewidth]{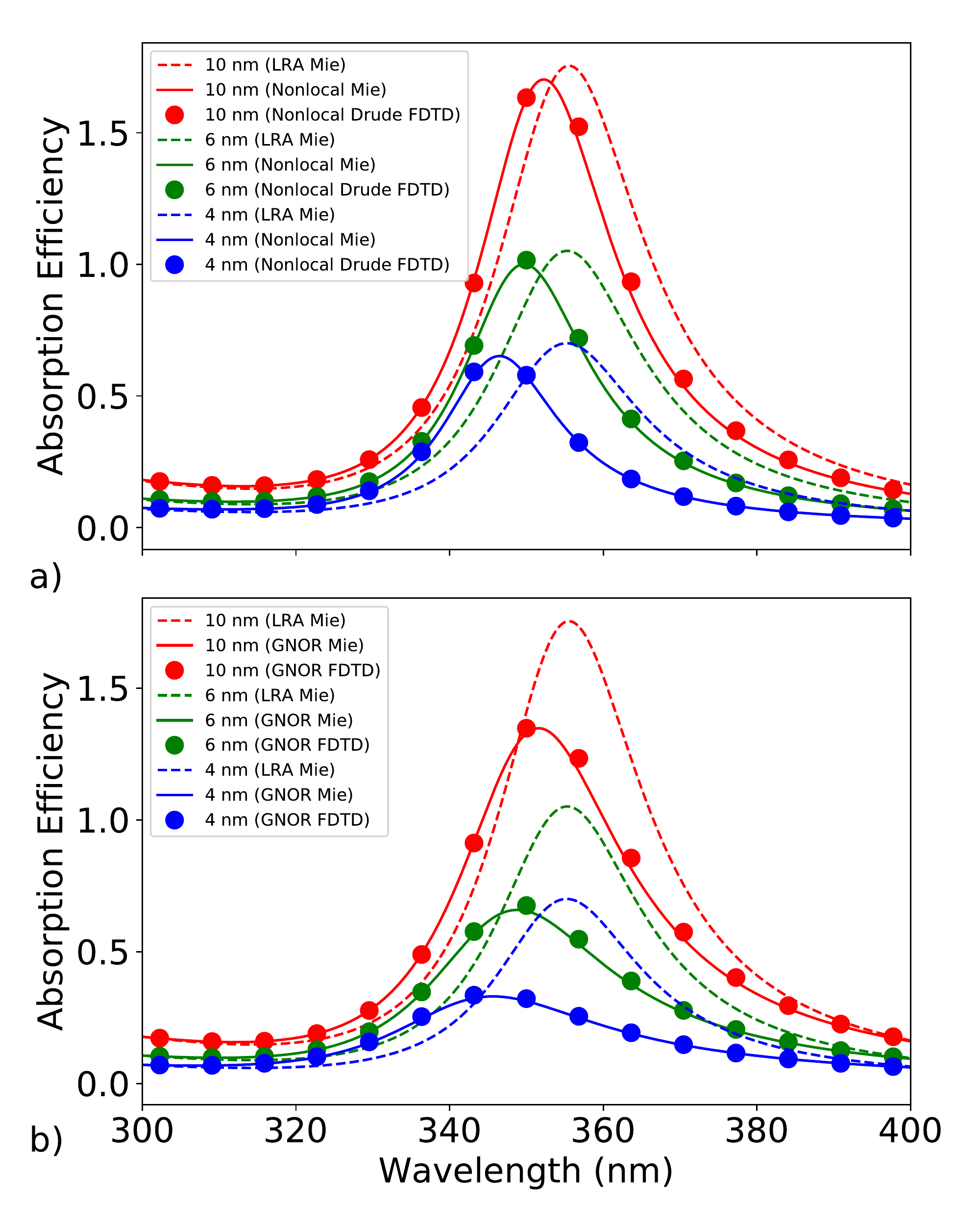}
\caption{Absorption cross sections for silver spheres of diameter 4 nm (blue), 6 nm (green) and 10 nm (red) using a) nonlocal Drude FDTD (filled circles), and b) GNOR FDTD (filled circles). The solid lines represent the nonlocal Mie theory solutions a) without diffusion and b) with diffusion. The dashed lines in both plots represent the LRA-Mie theory solutions.}
\label{fig:LocNonloc}
\end{figure}

In Fig. \ref{fig:LocNonloc}, we plot the absorption efficiencies for silver nanospheres in vacuum for three different diameters -- 4 nm (blue), 6 nm (green) and 10 nm (red) -- and different free charge polarization models. To model silver, we use the fitting parameters reported in Ref. \cite{vial_new_2011} for $\omega_p$ and $\gamma$ in Eqs. \ref{NonlocDrude} and \ref{GNOR}, and for all critical points model parameters in Eq. \ref{CritPoint}. We set $\beta^2=1/3v_F^2$ \cite{raza_nonlocal_2015} where $v_F=1.39\cdot 10^6$ m/s \cite{kittel_introduction_2005}. 

The FDTD domain is a 200 cell $\times$ 200 cell $\times$ 200 cell box truncated by a convolutional perfectly matched layer (CPML) \cite{taflove_computational_2005} consisting of 20 additional cells at each boundary. We use a uniform step-size of $\Delta x=\Delta y = \Delta z$ = $D$/100, where $D$ is the diameter of the sphere. For the particle sizes of interest here, this guarantees both that the step size is less than the Fermi wavelength, and that the spherical shape is sufficiently resolved. The total number of iterations vary with step-size and therefore particle size. For a 10 nm diameter particle, $3\cdot 10^5$ iterations are used to reach convergence; this is scaled appropriately for the other particle sizes.

In Fig. \ref{fig:LocNonloc} (a), we compare our nonlocal Drude FDTD calculations (filled circles) with classical (LRA) Mie theory solutions (dashed lines), and with nonlocal Mie theory solutions (solid lines). Nonlocal Mie theory \cite{ruppin_optical_1973,ruppin_optical_1975,pack_failure_2001} is a modified version of Mie theory \cite{hulst_light_1981} that allows for longitudinal modes, which permits a non-zero free charge density within the nanoparticle. It has been successfully validated with  experimental results for small particles larger than several nanometers. For example, a quasistatic version predicted absorption peaks that agree quantitatively with experiment for particle diameters down to 10 nm with a further qualitative agreement down to 2 nm \cite{raza_blueshift_2013-1}.

We find in Fig. \ref{fig:LocNonloc} (a) excellent agreement between nonlocal Drude FDTD and nonlocal Mie theory, with less than 2\% mean error for all three particle sizes. Doubling the FDTD step-size increases the mean error for the 10 nm particle case to 6.4\%. Halving the step-size decreases it to 1.3 \%. Thus reasonable convergence is reached with reasonable simulation domain sizes. As expected, we see an increasing blue-shift in the absorption peak with decreasing nanoparticle size with respect to the LRA Mie theory solution.

In Fig. \ref{fig:LocNonloc} (b) we compare our GNOR-FDTD calculations (filled circles) with LRA Mie theory solutions (dashed lines), and nonlocal diffusive Mie theory solutions (solid lines), where diffusion is accounted for by substituting
$\beta^2$ with $\beta^2 + \mathcal{D}\gamma - i\mathcal{D}\omega$ in nonlocal Mie theory \cite{raza_nonlocal_2015}. The free electron diffusion coefficient in silver is taken as $\mathcal{D}= 3.61\cdot 10^{-4}$ m\textsuperscript{2}/s \cite{raza_nonlocal_2015-2}. We again see excellent agreement, with less than 1.6\% mean error in GNOR-FDTD relative to nonlocal diffusive Mie theory for all three particle sizes. As expected the resonance positions predicted by the nonlocal Drude FDTD in Fig. \ref{fig:LocNonloc} (a) and GNOR FDTD in Fig. \ref{fig:LocNonloc} (b) are the same, with an increased line width for GNOR FDTD.


We now turn to examining the near field and free electron density distributions produced by nonlocal Drude FDTD, comparing to those produced by LRA FDTD.  In Fig. \ref{fig:FieldDist} we show the electric field amplitude distribution for the simulations of the 4 nm diameter silver nanoparticle produced by (a) LRA FDTD and (b) nonlocal Drude FDTD, for the wavelength corresponding to the peak of the (nonlocal) absorption spectrum ($\lambda=343$ nm). Shown are cuts in the $xz$ plane through the centre of the particle, where the incident plane wave is z-polarized and propagates along the y-axis. The field amplitudes are normalized, corresponding to an input field of 1 V/m. For higher quality images, we used a halved step size of $\Delta x$ = $D$/ 200. LRA FDTD produces a constant electric field within the sphere, as expected, while nonlocal Drude FDTD produces a field gradient due to the non-zero free charge distribution. 

\begin{figure}[tp!]
\centering
\includegraphics[width=\linewidth]{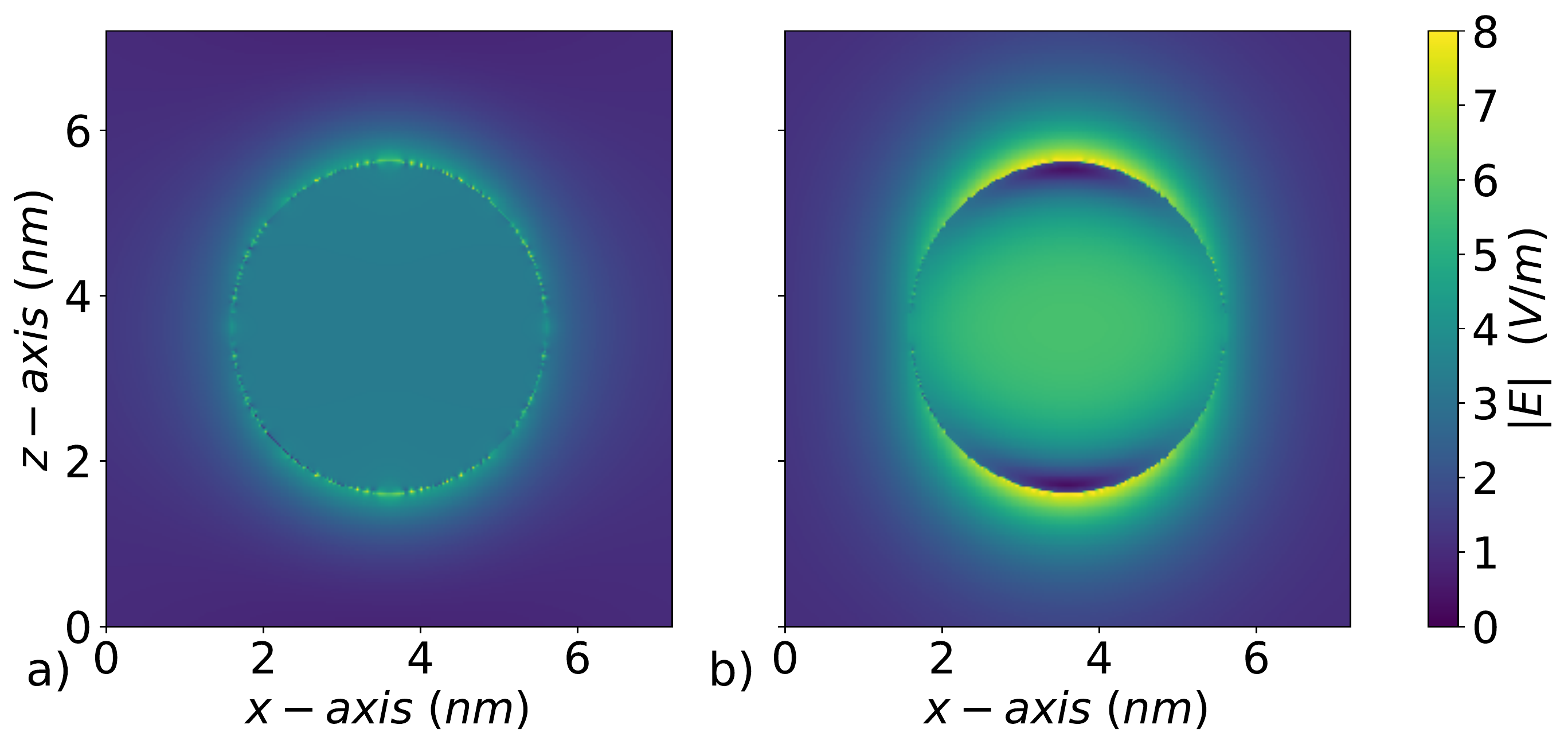}
\caption{Electric field amplitude distribution of a 4 nm diameter silver sphere in vacuum at $\lambda = 343$ nm using a) LRA FDTD and b) nonlocal Drude FDTD.}
\label{fig:FieldDist}
\end{figure}

\begin{figure}[tp!]
\centering
\includegraphics[width=\linewidth]{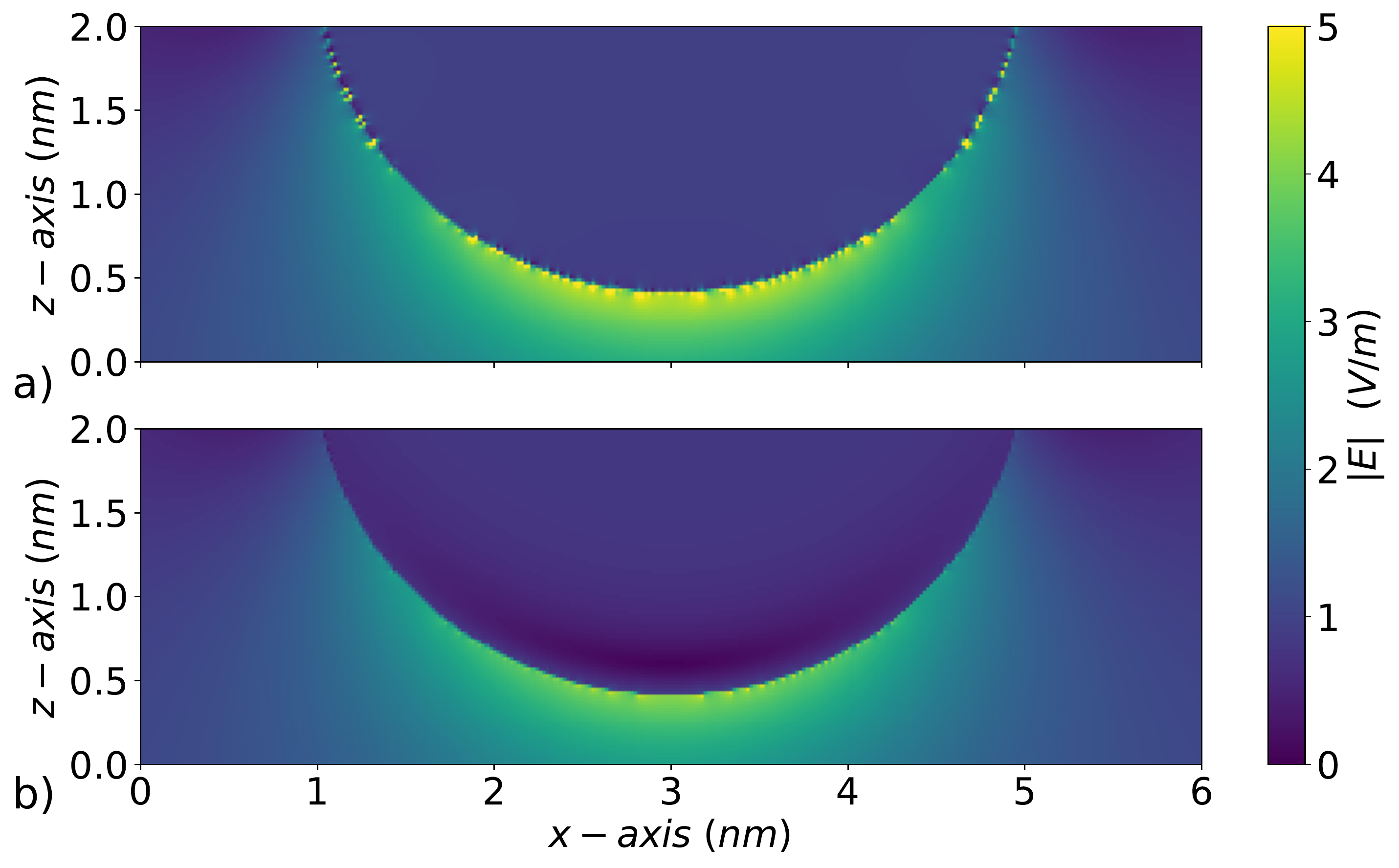}
\caption{Close-up of the electric field amplitude distribution of a 4 nm diameter silver sphere in vacuum at $\lambda = 425$ nm. Staircasing artifacts are much more prominent in a) for LRA FDTD than in b) for nonlocal Drude FDTD.}
\label{fig:SphereStair}
\end{figure}

Unexpected in Fig. \ref{fig:FieldDist} is the significant difference in the appearance of the fields at the particle boundary. While the effects of staircasing in LRA FDTD are clearly visible at the edges of the sphere in Fig. \ref{fig:FieldDist} (a), they appear smoothed out for nonlocal Drude FDTD in Fig. \ref{fig:FieldDist} (b). This is even more evident in Fig. \ref{fig:SphereStair} which shows the normalized electric field amplitude distribution produced by a) LRA FDTD and b) nonlocal Drude FDTD at  $\lambda = 425$ nm, a wavelength where the absorption efficiency and near fields for both approaches are almost identical, except for dramatic differences at the sphere boundary. The staircasing-induced boundary artifacts seen for LRA FDTD are clearly reduced with nonlocal Drude FDTD. 

Artifacts at the particle boundary are especially problematic for calculations that involve fields just outside the particle, such as, for example, determining the plasmonic near field enhancement of fluorescence, or engineering the spontaneous emission lifetime of fluorophosphores \cite{guzatov_plasmonic_2012}. This could also be important for calculations of plasmonics-enhanced nonlinear optics, where enhanced near fields close to plasmonic boundaries can be harnessed to not only enhance nonlinear optical processes by orders of magnitude, but also shape nonlinear optical fields \cite{li_nonlinear_2017,krasnok_nonlinear_2018,keren-zur_shaping_2018}. With nonlocal FDTD, the reduction of staircasing artifacts would result in more reliable calculations.


In Fig. \ref{fig:EdensDist} we plot the free electron density distribution corresponding to the 4 nm silver particle simulations of Fig. \ref{fig:FieldDist}. Shown are cuts in the $xz$ plane through the center of the sphere for $\lambda$ = 343 nm, as produced by (a) LRA FDTD and (b) nonlocal Drude FDTD. As expected, nonlocal Drude FDTD allows for the spread of the charge distribution near the particle boundary, which we see is on the order of the Fermi wavelength for silver, $\lambda_F$ = 0.5 nm. In contrast, for LRA FDTD the charge is bound to the surface.

\begin{figure}[tp!]
\centering
\includegraphics[width=\linewidth]{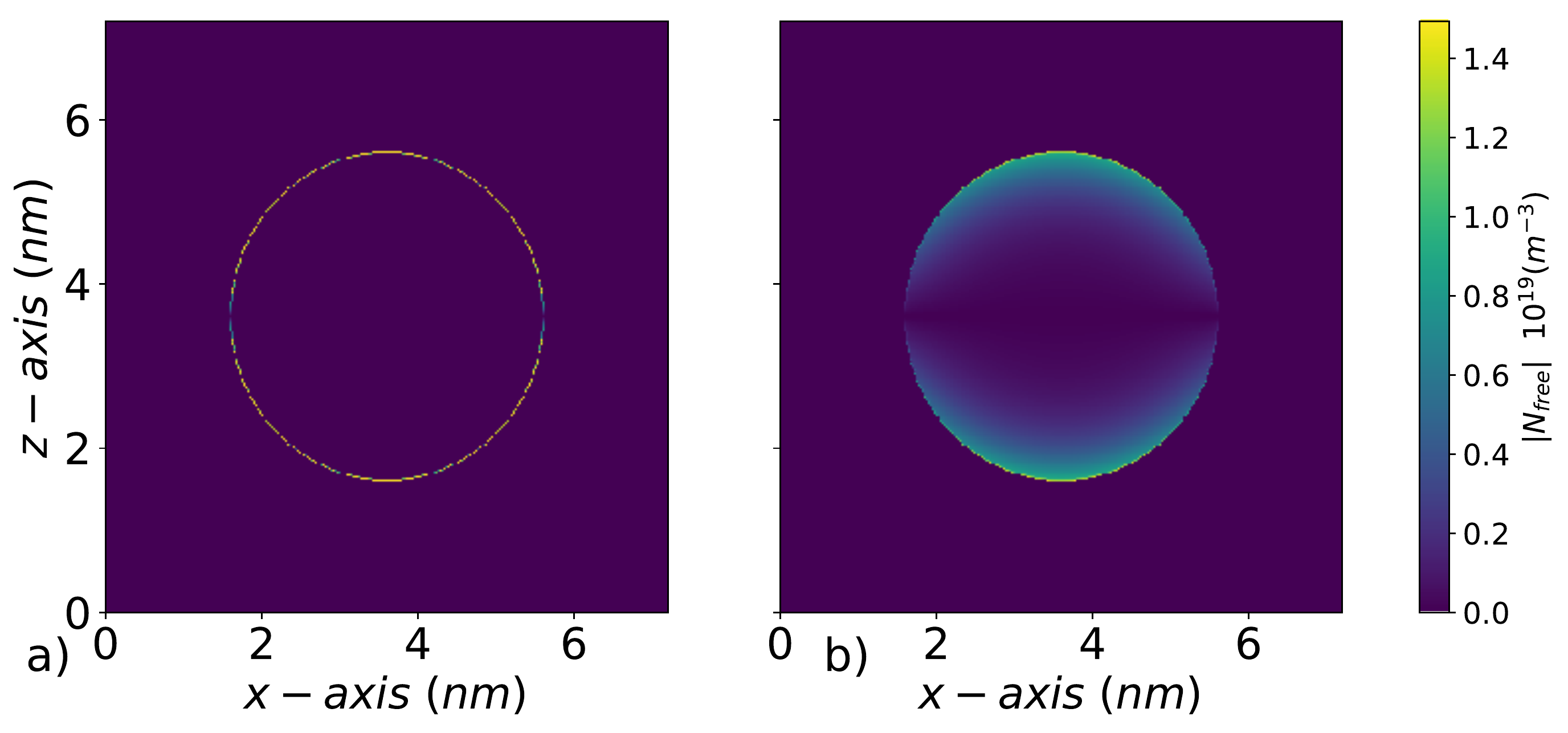}
\caption{Free electron density within a 4 nm diameter silver sphere in vacuum at $\lambda =$ 343 nm using a) LRA FDTD and b) nonlocal Drude FDTD.}
\label{fig:EdensDist}
\end{figure}


We present in the Supplementary Information (SI) time-domain movies for the 4 nm silver sphere simulations corresponding to Figs. \ref{fig:FieldDist} - \ref{fig:EdensDist} where the incident plane wave pulse function is a normalized raised cosine $f(t)=[(1-\cos(\omega_{max} t))/2]^3$, where $\omega_{max} = 1.26\cdot 10^{16}$ rad/s is the maximum frequency of interest (corresponding to $\lambda = $ 150 nm). Included in the SI are movies of the electric field amplitude dynamics for nonlocal Drude FDTD in the $xz$ and $yz$ planes, \textit{movie1}, and \textit{movie2}, respectively, where both planes cut through the center of the particle. The comparable movies for LRA FDTD are \textit{movie3} and \textit{movie4}. The nonlocal FDTD movies show a radially propagating wave inside the sphere whereas the LRA FDTD movies do not; in the latter, the field is almost always constant across the sphere (as expected). Further, while staircasing artifacts in the LRA FDTD movies are quite pronounced, they are hardly visible in the nonlocal Drude FDTD movies. The corresponding free electron density movie for nonlocal Drude FDTD is \textit{movie5}. Since the free electron density does not need to be tracked within the LRA FDTD implementation, and all charge is strictly localized to the surface, we do not include a time-domain movie for this case. 



%


We now turn to further validate our FDTD implementations by comparing our simulated results to those of experiment. In Ref. \cite{scholl_quantum_2012}, electron energy loss spectroscopy measurements are presented for silver nanosphere diameters ranging from 2 to 24 nm on a carbon film and compared to Mie theory calculations that use size-dependent permittivities derived quantum mechanically. Using GNOR FDTD, we calculate the absorption spectra of silver nanospheres in a $n$ = 1.3 dielectric background (as used in the calculations of Ref. \cite{scholl_quantum_2012}) with diameters ranging from 2 to 24 nm, and plot these in Fig. \ref{Spheretrans}. We find the same trend as presented in Ref.\cite{scholl_quantum_2012}, showing that GNOR FDTD is consistent with experimental measurements and calculations using quantum-based permittivities, with quantitative agreement down to 10 nm diameter, and qualitative agreement to 2 nm. As discussed in Ref. \cite{raza_nonlocal_2015}, for diameters less than 10 nm, the nonlocal model predicts a resonance shift that is not as large as determined by experimental measurements, which is consistent with our results. This may be due to more complicated phenomena occurring in silver, such as inhomogeneous equilibrium electron density, or spill-out effects, that are not included in the GNOR model we have implemented.


\begin{figure}[tp!]
\centering
\includegraphics[width=3.7in]{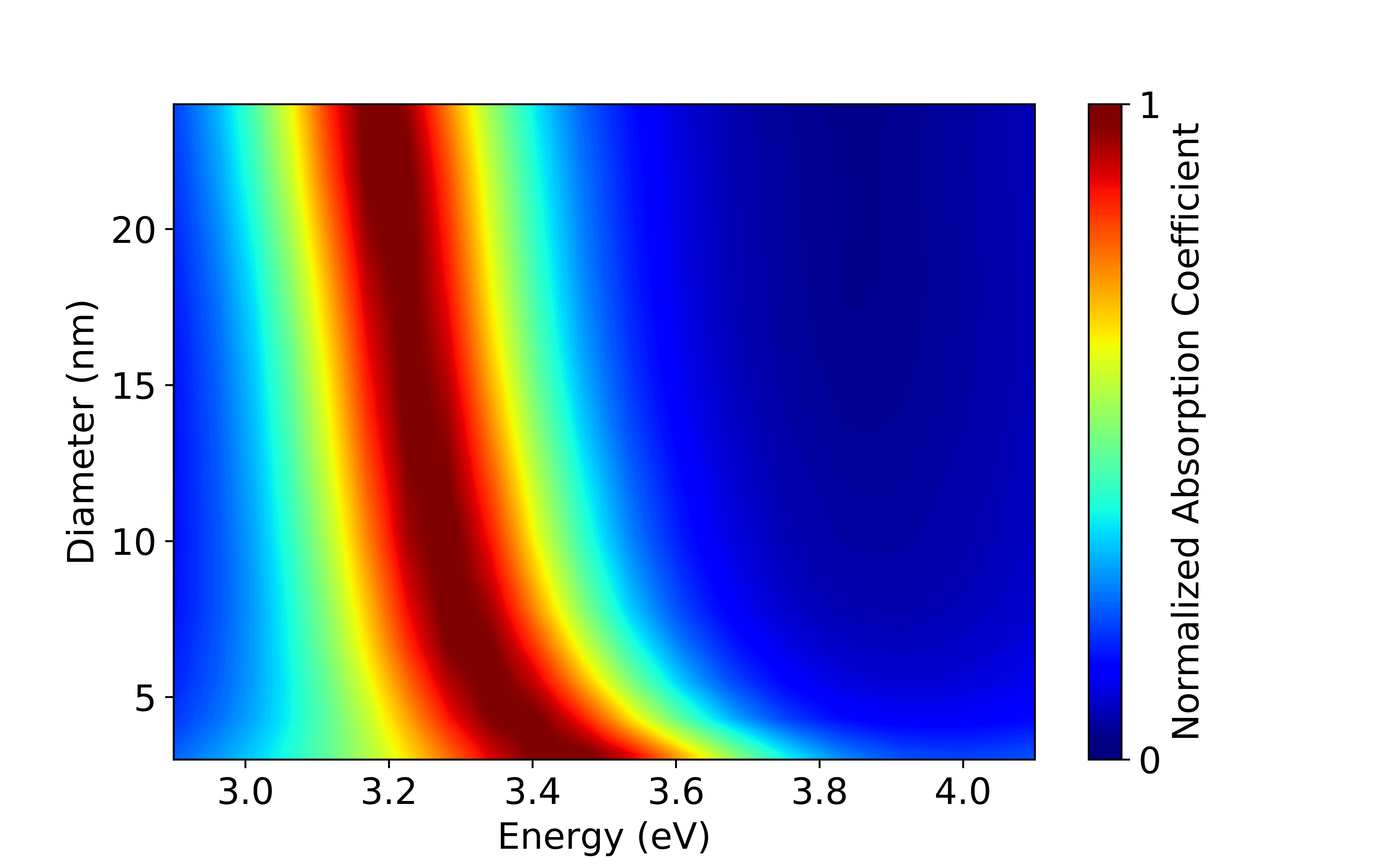}
\caption{Contour plot of the absorption cross section of silver spheres embedded in a $n$ = 1.3 dielectric background as a function of incident plane wave photon energy (horizontal axis) and sphere diameter (vertical axis).}
\label{Spheretrans}
\end{figure}

%



\begin{figure}[tp!]
\centering
\includegraphics[width=3in]{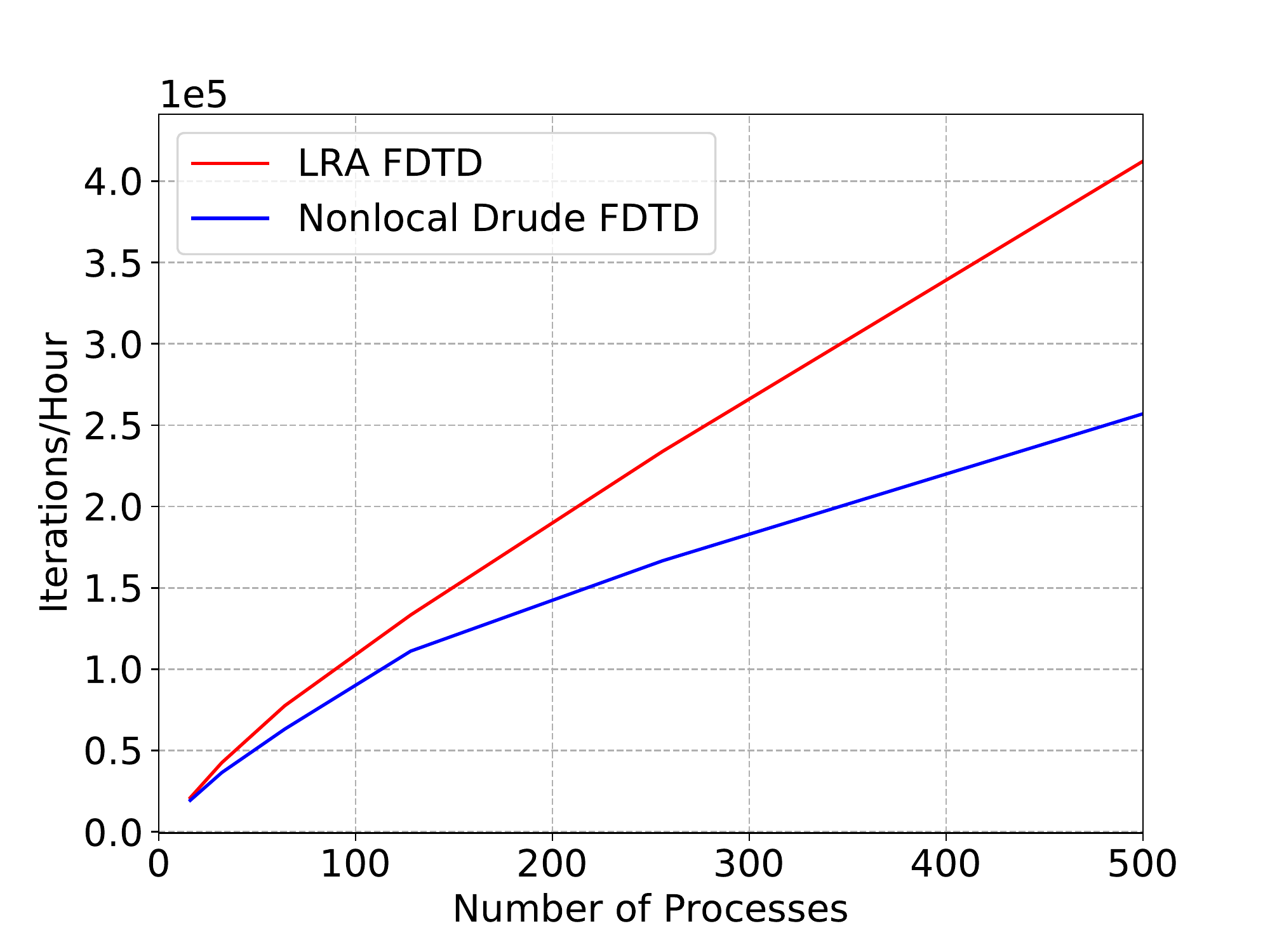}
\caption{Scalability of the nonlocal Drude FDTD (blue line) compared to that of LRA FDTD (red line). Plotted is the number of FDTD time iterations computed per hour versus the number of processes used.}
\label{fig:Scalability}
\end{figure}

Finally, we turn to a discussion of computational resources and scalability. To give an example, for the 10 nm, 6 nm and 4 nm sphere simulations presented in Fig. \ref{fig:LocNonloc}, we used 64, 128, and 256 cores, respectively, on the Graham cluster operated by Compute Canada \cite{noauthor_graham_nodate}. Recall that the workload is heavier for the smallest particle since our step size $\Delta x$ (and thus time step size) is proportional to sphere diameter, and thus more iterations in time are required to reach convergence for the smallest particles. 

In general, when running simulations on a large number of cores, as we do for this paper, it is imperative to investigate the implementation's scalability, that is, the performance enhancement obtained by  increasing the number of CPUs. In the ideal case, the scaling is linear, meaning when one doubles the number of processes, the computation time is halved; this is often not achieved due to overhead, including inter-processor communications.


In Fig. \ref{fig:Scalability}, we plot the number of FDTD time iterations calculated per hour as a function of the number of processes used in the nonlocal Drude FDTD (blue line) and LRA FDTD (red line) simulations. The simulation set up is the same as for Fig. \ref{fig:LocNonloc}, where we vary the number of processes while keeping the total number of cells constant. For $\leq 128$ processes, the scalability of nonlocal Drude FDTD is nearly linear and comparable to LRA FDTD. While the scalability of LRA FDTD remains mostly linear for a larger number of processes, that of nonlocal Drude FDTD does not, likely due to the doubled inter-process communication required, as discussed in Section \ref{sec4}. As one increases the number of processes while maintaining the same domain size, the subdomain surface-to-volume ratio increases, and therefore the inter-process communication time eventually becomes comparable to the computation time within a time step. Thus, if we were to consider a larger domain size, we expect the near-linear scalability to extend to larger process numbers. 


\section{Nonlocal FDTD applied to larger, complex nanoparticles}\label{sec6}



In this section, we demonstrate the benefit of our parallel FDTD implementation even further, by considering a much larger plasmonic structure containing small, sharp features. Such a structure would still be expected to exhibit nonlocal effects, and thus simulating it via FDTD would still require a small grid cell size, small enough to resolve electron dynamics within the sharp features. We consider as an example a structure inspired by the star-shaped nanoparticles synthesized for Ref. \cite{minati_one-step_2014}, which are spherical particles around 50 nm in diameter with small 5 nm extrusions from their surfaces that come to a sharp tip.

The structure we simulate is a silver nanosphere with triangular nanoprisms extruding from the equator in the $xz$ plane, perpendicular to the incident plane wave propagation axis, as illustrated in Fig. \ref{SunCS}. The sphere is 40 nm in diameter and the nanoprisms are embedded into the sphere so that the effective prism length is 5 nm. The tips of the nanoprisms are rounded with a radius of 1 nm. The FDTD step-size is uniform with $\Delta x$ = 0.1 nm and the domain size is 600 $\times$ 600 $\times$ 600 Yee cells (not including CPMLs). The simulations are run for $4\cdot 10^5$ iterations over 1000 processes (10 $\times$ 10 $\times$ 10).


\begin{figure}[tp!]
\centering
\includegraphics[width=2in]{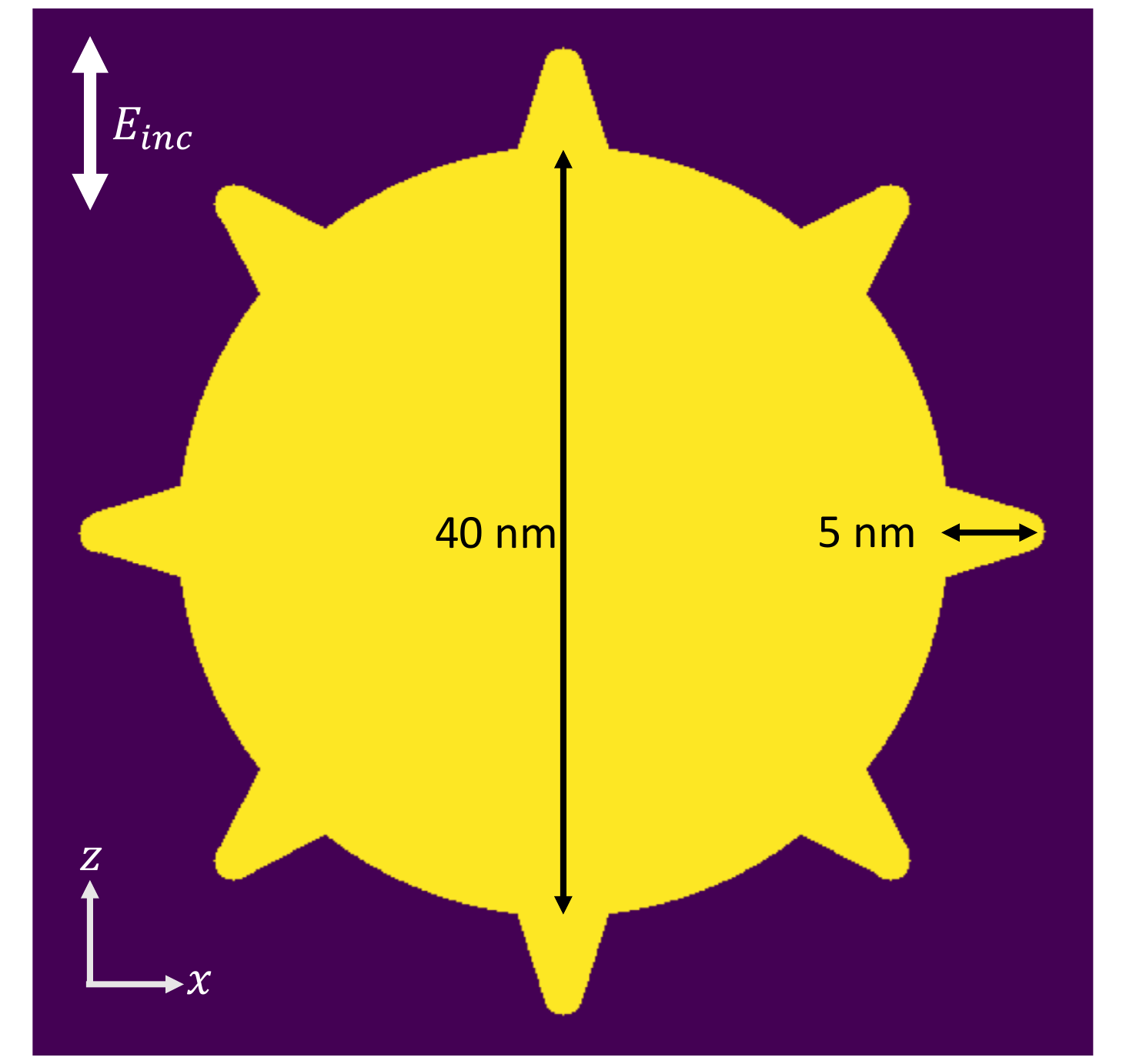}
\caption{Cross section in the $xz$ plane of the star-shaped silver nanoparticle, consisting of a sphere with 40 nm diameter and triangular nanoprism extrusions extending 5 nm from the sphere surface. The incident plane wave used to illuminate this structure is z-polarized and propagates along the y axis.}
\label{SunCS}
\end{figure}

The absorption spectrum of this particle is shown in Fig. \ref{SunAbs}, calculated using both LRA FDTD (red) and nonlocal Drude FDTD (blue). There are two resonances attributed to the star-shaped particle. The peaks near 355 nm corresponds to the plasmonic resonance of the sphere, and these completely overlap for the two models. This is expected, as nonlocal effects should be negligible for particles larger than 20 nm. The peaks near 575 nm, however, correspond to the resonance of the nanoprisms, specifically the ones aligned parallel to the incident field polarization. Here we do see that nonlocal effects become important, as the peak wavelength predicted by nonlocal Drude FDTD is 10 nm blue shifted from that predicted by LRA FDTD.

\begin{figure}[tp!]
\includegraphics[width=3in]{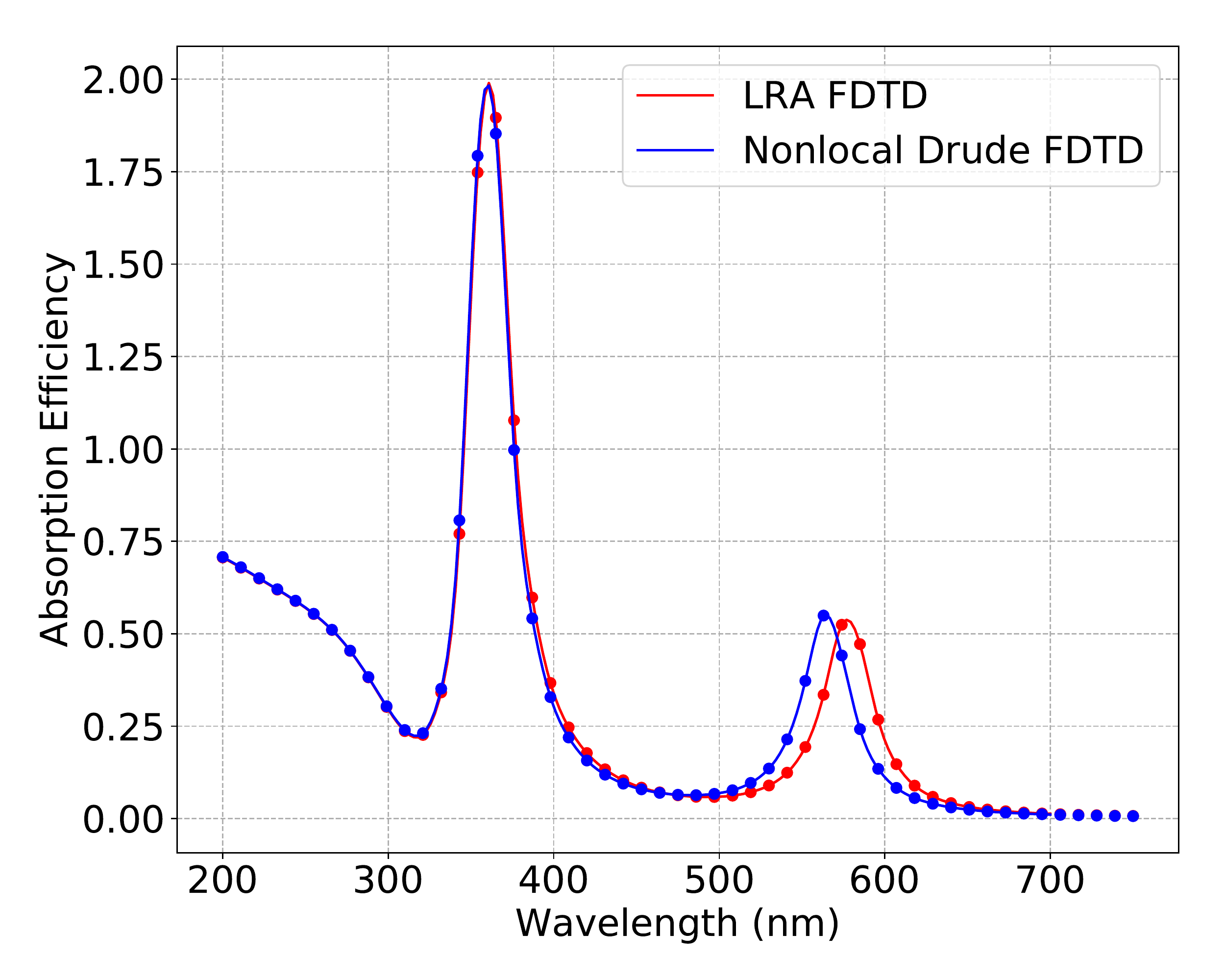}
\caption{Absorption efficiency of the silver star-shaped nanoparticle calculated using LRA FDTD (red dots) and nonlocal Drude FDTD (blue dots); the lines are interpolations of the FDTD data.}
\label{SunAbs}
\end{figure}

Additional differences between the two models manifest in the field amplitude distributions for wavelengths below the interband transition wavelength $\lambda_{IB}\approx$ 330 nm. The electric field amplitude distributions at $\lambda$ = 320 nm are shown in Fig. \ref{StarFieldDist} for (a) LRA FDTD and (b) nonlocal Drude FDTD. Standing waves inside the vertical triangles, with a wavevector in the z-direction, are visible in Fig. \ref{StarFieldDist} (b) for nonlocal Drude FDTD.
This is a longitudinal mode\cite{raza_unusual_2011}, and is expected since its wavelength lies below the epsilon-near-zero wavelength, which in silver is approximately equal to $\lambda_{IB}$. In the LRA, these modes cannot be excited by an incident transverse wave, and we see in Fig. \ref{StarFieldDist} (a) that they are not. However, with nonlocality, such excitation can occur \cite{raza_unusual_2011} due to the Pekar additional material boundary conditions, discussed in Section \ref{sec3}.

\begin{figure}[tp!]
\centering
\includegraphics[width=\linewidth]{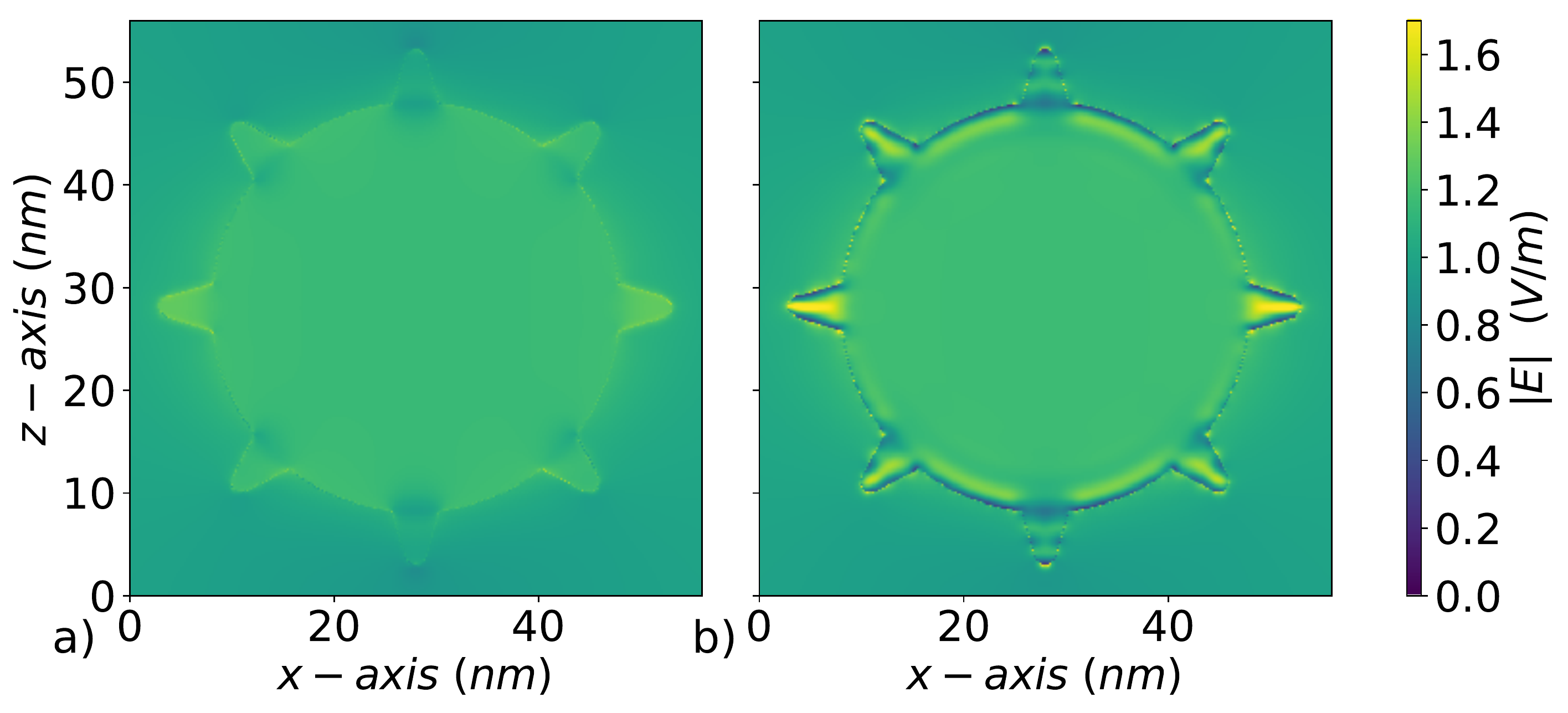}
\caption{Electric field amplitude distribution of a star-shaped nanoparticle in vacuum at $\lambda=320$ nm using a) LRA FDTD and b)  nonlocal Drude FDTD.}
\label{StarFieldDist}
\end{figure}

These standing waves become more pronounced for isolated nano-triangles as shown in Fig. \ref{TriFieldDist} (b) where nonlocal Drude FDTD was used, whereas they are absent in Fig. \ref{TriFieldDist} (a), where LRA FDTD was used. The standing waves found for nonlocal FDTD are damped by interband transitions for wavelengths below 300 nm.

We now turn to examining the effect of staircasing for the star-shaped nanoparticle. We show in Fig. \ref{Star-Staircasing} a zoomed-in view of the electric field amplitude distribution at $\lambda = $ 475 nm, a wavelength where there is more contrast at the particle boundary than for the wavelength considered in Fig. \ref{StarFieldDist}. Despite a very fine mesh of $\Delta x$ = 0.1 nm, one can clearly see staircasing artifacts in the electric field amplitude distribution for LRA FDTD in Fig. \ref{StarFieldDist} (a). These are notably reduced in Fig. \ref{StarFieldDist} (b), where nonlocal Drude FDTD was used. 


This reduction in staircasing artifacts is further illustrated by the time-domain movies of the star-shaped nanoparticle simulations presented in the SI. These depict the time evolution of the electric field amplitude and free electron density in the $xz$ and $yz$ planes through the center of the particle, where the incident plane wave is polarized in $z$ and propagates along $y$. The electric field evolution movies for nonlocal Drude FDTD are \textit{movie6} and \textit{movie7}, while those for the corresponding free electron density evolution are \textit{movie8} and \textit{movie9}. The electric field evolution movies for LRA FDTD are \textit{movie10} and \textit{movie11}.


\begin{figure}[tp!]
\centering
\includegraphics[width=\linewidth]{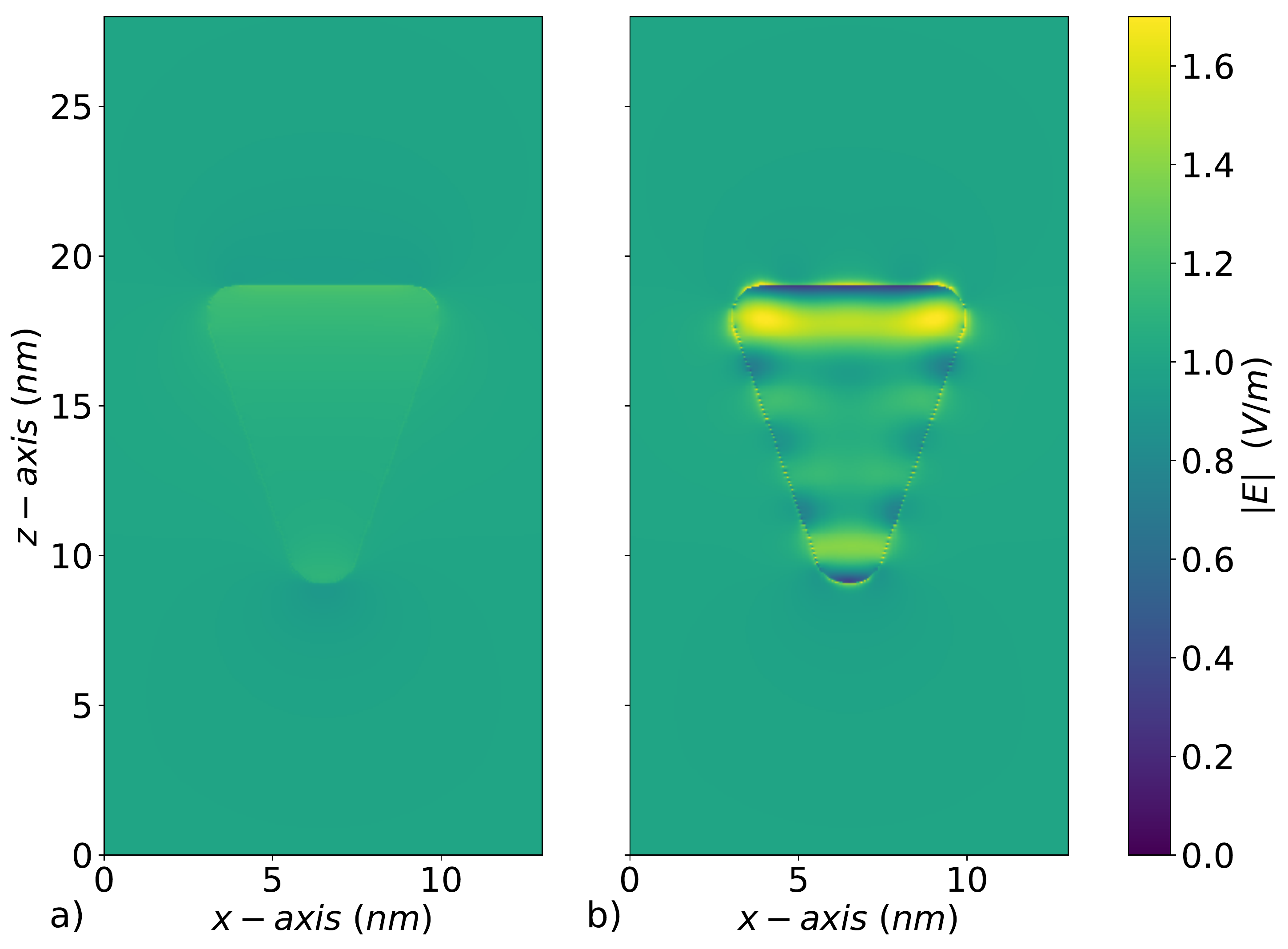}
\caption{Electric field amplitude distribution of isolated nanotriangles in vacuum at $\lambda=320$ nm using a) LRA FDTD and b) the nonlocal Drude FDTD.}
\label{TriFieldDist}
\end{figure}

\begin{figure}[tp!]
\centering
\includegraphics[width=\linewidth]{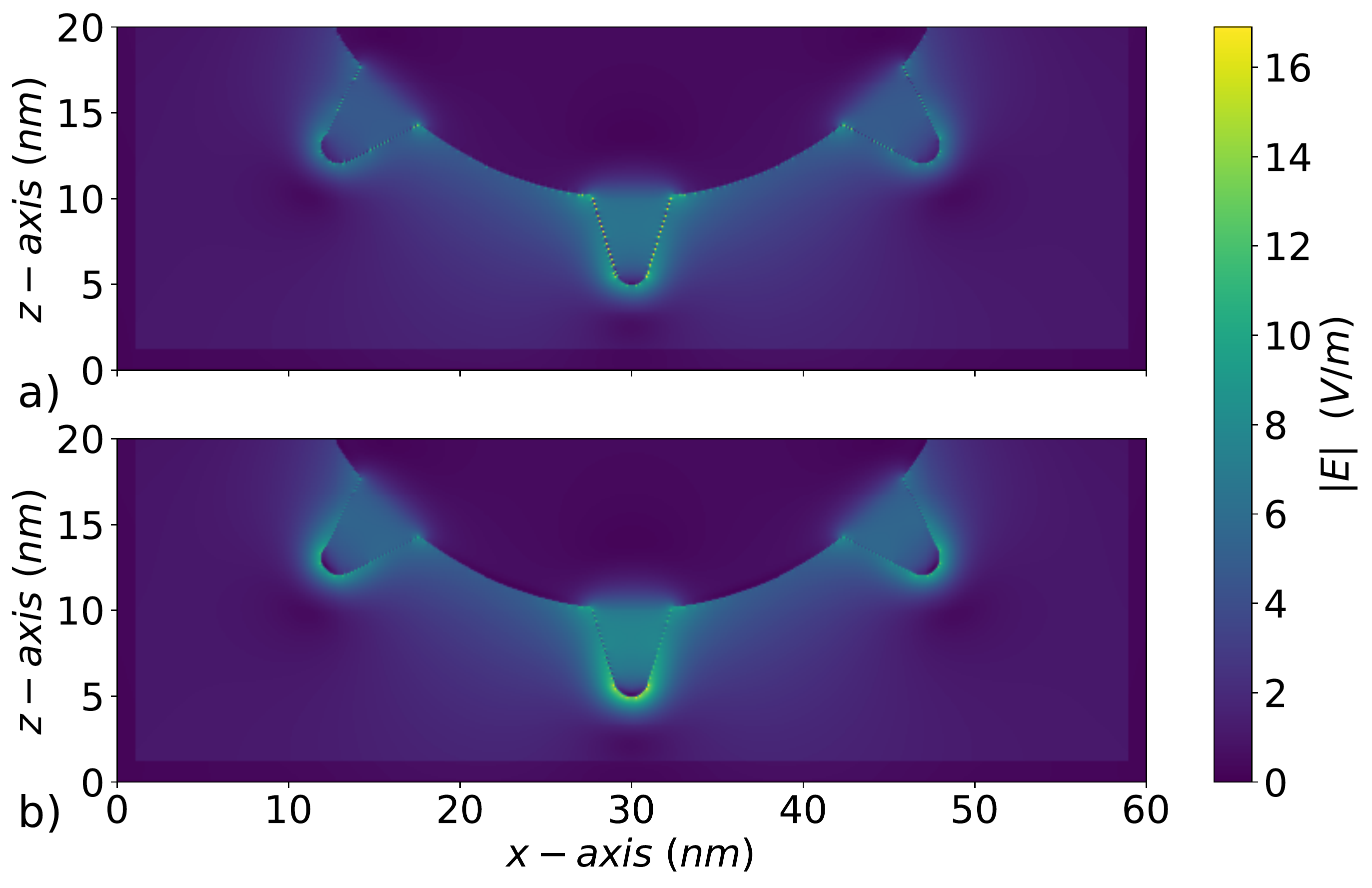}
\caption{Close up of the electric field distribution of the star-shaped particle at $\lambda = $ 475 nm. The staircasing artifacts are much more prominent in a) for LRA FDTD than in b) for nonlocal Drude FDTD.}
\label{Star-Staircasing}
\end{figure}


\section{Conclusion} \label{sec7}
We have introduced a parallel FDTD implementation for modelling nonlocality in plasmonics. We used an auxiliary differential equation approach to model nonlocality with and without electron diffusion, and described in detail how to use the message passing interface framework for parallel computation via domain decomposition. After validating our implementation via comparisons with analytical and experimental results for small nanospheres, we demonstrated the robustness of our parallel implementation for larger particles with sharp nanoscale features. We find that the inclusion of nonlocality within FDTD significantly reduces staircasing artifacts that can plague standard plasmonic FDTD modelling based on the local response approximation (LRA). This suggests that beyond its importance for modelling nonlocality, nonlocal FDTD might be advantageous for calculations that require precise values for the fields at plasmonic boundaries. This includes, for example, determining plasmonic fluorescence enhancement, plasmonics-mediated fluorescent lifetime engineering, and plasmonics enhanced nonlinear optics.

\section*{Acknowledgment}
The authors would like to thank Compute Canada and Scinet for computational resources, SOSCIP for their computational resources and financial support, and the Natural Sciences and Engineering Research Council of Canada and the Canada Research Chairs program for financial support.

\ifCLASSOPTIONcaptionsoff
  \newpage
\fi



\bibliographystyle{IEEEtran}
\bibliography{Nonlocality}

%

%



\begin{IEEEbiography}[{\includegraphics[width=1in,height=1.25in,clip,keepaspectratio]{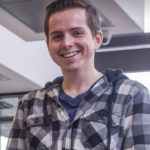}}]{Joshua Baxter} earned his BSc in Physics/Mathematics from the University of Ottawa in 2017 and started his MSc in the same year. He transferred to his PhD in 2019. His work has involved plasmonic colouring with ultrafast lasers, simulations of laser heating, and computational nanophotonics. His current work focuses on Deep Learning applications to plasmonic colouring and nonlinear plasmonics through hydrodynamics. 
\end{IEEEbiography}

\begin{IEEEbiographynophoto}{Antonino Cal\`a Lesina}
is a Research Associate at the University of Ottawa, Canada, since 2018. He obtained his B.Sc. in Electronics Engineering in 2006 and his M.Sc. in Telecommunications Engineering in 2009 from the University of Catania, Italy, and his Ph.D. in Information and Communication Technologies in 2013 from the University of Trento, Italy, conducting his research in computational electromagnetics at the Fondazione Bruno Kessler. He joined the University of Ottawa in 2013 as a Postdoctoral Fellow in the groups of professors Lora Ramunno and Pierre Berini. His research focuses on computational nanophotonics, nonlinear and tunable metasurfaces, plasmonics and metamaterials.
\end{IEEEbiographynophoto}


\begin{IEEEbiographynophoto}{Lora Ramunno}
obtained her Ph.D. in physics from the University of Toronto in 2002. After a year in a high-tech start-up, she returned to academia as a postdoctoral fellow at uOttawa. Since 2007, she has been a faculty member in Physics at uOttawa where she is currently Full Professor. She held a Canada Research Chair in Computational Nanophotonics from 2007-2017, and was the recipient of an Early Researcher Award in 2007. Her theoretical/computational research focuses on nonlinear and nano-optics, including intense laser-matter interaction, nanophotonics,  nonlinear optical microscopy and photonic devices. She also has several successful industrial collaborative projects. Prof. Ramunno is a fellow of the Max-Planck-UOttawa Centre for Extreme and Quantum Photonics, a fellow of the NRC-UOttawa Joint Centre for Extreme Photonics, and a member of the uOttawa Centre for Research in Photonics.
\end{IEEEbiographynophoto}




\end{document}